\documentclass[aps,physrev,groupedaddress]{revtex4-2}

\usepackage[utf8]{inputenc}
\usepackage[margin=1.0in]{geometry}
\usepackage[draft,authormarkup=none]{changes}
\usepackage{amsmath}
\usepackage{xcolor}
\usepackage{caption}
\usepackage{cancel}
\usepackage{subcaption}
\usepackage{graphicx}
\usepackage[citecolor=blue]{hyperref}
\usepackage{overpic}
\usepackage[usestackEOL]{stackengine}
\usepackage{multirow}
\usepackage{placeins}

 \hypersetup{colorlinks=true,linkcolor=blue,filecolor=magenta,urlcolor=blue,
     }

\newcommand{\urms}{u_\text{RMS}}

\newcommand{\abbrev}[1]{\mathrm{#1}}
\newcommand{\NMSE}{\abbrev{NMSE}}
\newcommand{\MAX}{\abbrev{MAX}}

\newcommand{\PSNR}{\abbrev{PSNR}}
\newcommand{\SSIM}{\abbrev{SSIM}}

\usepackage{tikz}
\usetikzlibrary{positioning}
\usetikzlibrary{backgrounds, shadows.blur}
\newcommand{\labelbox}[3]{%
  \node[fill=white, fill opacity=0.8, text opacity=1,
        rounded corners=1pt, inner sep=3pt,
        font=\normalsize, #1] at (#2) {#3};
}

\begin{document}
\preprint{APS}

\newif\ifshowedits
\newif\ifshowdeletions

\showeditstrue 
\showdeletionstrue
\showeditsfalse
\showdeletionsfalse

\ifshowedits
\definechangesauthor[name=kristoffer]{kristoffer}
\definechangesauthor[name=jorgen]{jorgen}
\definechangesauthor[name=simen]{simen}
\definechangesauthor[name=nathan]{nathan}
\newcommand{\delete}[1]{{\color{gray}\sout{#1}}}
\newcommand{\add}[1]{{\color{blue}#1}}
\newcommand{\replace}[2]{\delete{#1}\add{#2}}

\newcommand{\akr}[1]{\added[id=kristoffer]{#1}}
\newcommand{\dkr}[1]{\delete{#1}}
\newcommand{\rkr}[2]{\replace{#1}{#2}}

\newcommand{\ajo}[1]{\added[id=jorgen]{#1}}
\newcommand{\djo}[1]{\delete{#1}}
\newcommand{\rjo}[2]{\replace{#1}{#2}}

\newcommand{\asm}[1]{\added[id=simen]{#1}}
\newcommand{\dsm}[1]{\delete{#1}}
\newcommand{\rsm}[2]{\replace{#1}{#2}}

\newcommand{\ana}[1]{\added[id=nathan]{#1}}
\newcommand{\dna}[1]{\delete{#1}}
\newcommand{\rna}[2]{\replace{#1}{#2}}
\else
\newcommand{\delete}[1]{}
\newcommand{\add}[1]{#1}
\newcommand{\replace}[2]{\add{#2}}
\newcommand{\akr}[1]{#1}    \newcommand{\dkr}[1]{}    \newcommand{\rkr}[2]{#2}
\newcommand{\ajo}[1]{#1}    \newcommand{\djo}[1]{}    \newcommand{\rjo}[2]{#2}
\newcommand{\asm}[1]{#1}    \newcommand{\dsm}[1]{}    \newcommand{\rsm}[2]{#2}
\newcommand{\ana}[1]{#1}    \newcommand{\dna}[1]{}    \newcommand{\rna}[2]{#2}

\fi
\ifshowdeletions \else \renewcommand{\delete}[1]{}
\fi

\title{Mapping surface height dynamics to subsurface flow physics in free-surface turbulent flow using a shallow recurrent decoder}
\author{Kristoffer S. Moen$^1$}\thanks{K.S.M.\ and J.R.Aa.\ are to be considered joint first authors.}
\author{J{\o}rgen R. Aarnes$^1$$^*$}
\author{Simen {\AA}. Ellingsen$^1$}
\author{J. Nathan Kutz$^{2,3}$}
\affiliation{$^1$Norwegian University of Science and Technology, Trondheim, Norway}
\affiliation{$^2$Department of Applied Mathematics, University of Washington, Seattle, Washington, USA}
\affiliation{$^3$Autodesk Research, 6 Agar Street, London, United Kingdom}

\begin{abstract}
Near-surface turbulent flows beneath a free surface are reconstructed from sparse measurements of the surface height variation, by a novel neural network algorithm known as the {\em SHallow REcurrent Decoder} (SHRED). 
The reconstruction of turbulent flow fields from limited, partial, or indirect measurements remains a grand challenge in science and engineering. The central goal in such applications is to leverage easy-to-measure proxy variables in order to estimate 
quantities which have not been, and perhaps cannot in practice be, measured. 
\djo{Specifically, i}\ajo{I}n the application considered here, the aim is to use a sparse number of surface height point measurements of a flow field, or drone video footage of surface features, in order to infer the turbulent flow field beneath the surface.  
SHRED is a deep learning architecture that learns a delay-coordinate embedding from a few surface height (point) sensors and maps it, via a shallow decoder trained in a compressed basis, to full subsurface fields, enabling fast, robust training from minimal data. 
We demonstrate the SHRED sensing architecture on%
\replace{both }{
two types of turbulent data from recent studies (Aarnes \emph{et al.} J.~Fluid Mech.\ \textbf{1007} A38, 2025 and Babiker \emph{et al.} Phys.\ Rev.\ Fluids \textbf{11} 054802, 2026, respectively):
}%
fully resolved DNS data and PIV laboratory data from a turbulent water tank. SHRED is capable of robustly mapping surface height fluctuations to full-state flow fields up to about \rkr{two integral length scales}{one integral length scale} deep, with as few as three surface measurements.   
\end{abstract}

\maketitle

\newcommand{\bS}{\mathbf{S}}
\newcommand{\bU}{\mathbf{U}}
\newcommand{\bV}{\mathbf{V}}
\newcommand{\bW}{\mathbf{W}}

\section{Introduction}

The surface of a gently flowing river is not flat, but comprises of features that are characteristic imprints of the turbulent flow beneath it. A qualitative taxonomy groups the coherent, long-lived structures into `dimples', `boils', and `scars' \cite{brocchini2001}. These structures are imprints of surface-attached vortices, upwelling events, and strong surface-tangential vortices \cite{aarnes2025}, respectively. Recent investigations have shown that features of the turbulent free surface elevation which are easily distinguished with computer vision are closely correlated to the turbulent velocity field \cite{babiker2023}, which is otherwise impractical to measure outside the laboratory. Obtaining estimates of the turbulent velocity field at or close beneath the free surface is of high practical interest. For instance, near-surface turbulence controls the rate at which heat and gas are transferred between water and air \cite{wanninkhof2009, dasaro2014}, and the total greenhouse gas discharge from rivers, where these characteristic imprints may be seen, is similar in magnitude to the total gas flux through the ocean surface \cite{brinkerhoff2022}. In-situ measurements are slow, expensive, and provide data for a single point or trajectory at a time, whereas observations of the surface, for instance using airborne drones, can cover larger areas comparatively quickly and at low cost. Such a technique, however, is still at a conceptual stage \cite{dolcetti2022} and will remain so until quantitative predictions of subsurface flow, and concomitant processes such as gas flux, can be made from surface observations only. We \rjo{propose}{present} a neural network scheme to directly map surface height variations to the underlying flow fields beneath the surface, thus helping close the gap in practice for using drone sensing.

Data-driven methods have gained traction in turbulence research. Applications include spatial super-resolution or reconstruction from sparse measurements, using shallow networks \cite{Erichson2020},  convolutional neural networks (CNNs) \cite{Zhang2023}, \akr{diffusion models} \cite{pant2021}, and generative adversarial networks (GANs) \cite{Deng2019_GANs}, as well as physics-informed neural networks (PINNs) as means to reduce non-physical deviations as accumulated by the preceding networks \cite{discetti2022, morenosoto2024}. Furthermore, temporal enhancement of sparse data 
\rjo{or}{and} reconstruction of turbulence has been attempted using shallow multilayer perceptron (MLP) \cite{morenosoto2024}, recurrent neural networks (RNN) \cite{Jin2020}, and long short-term memory networks (LSTM)\cite{Deng2019}. In recent years, RNNs, and particularly LSTM networks, have shown strong performance in handling the nonlinear time dynamics of turbulent flows \cite{Deng2019, BORRELLI2022109010,BruntonKutz2019, Srinivasan2019, morenosoto2024, discetti2022}.
LSTM networks are a subclass of RNN, designed to learn long-range dependencies in time series, and improves upon instabilities that may occur in general RNNs due to vanishing gradients in the training optimization.
The resulting nonlinear network is powerful in handling sequential datasets, such as time series, and particularly useful for capturing or forecasting nonlinear temporal dynamics \cite{BruntonKutz2019}. Traditional linear methods such as Proper Orthogonal Decomposition (POD), Linear Stochastic Estimation (LSE), and linear neural networks are far outperformed by nonlinear networks in most turbulence cases, as pointed out in several papers over the last few decades \cite{Brunton2020, Cuellar2024, Milano,  xuan2023,discetti2022}. Nonlinear methods are more generalizable across different turbulent cases, whereas traditional linear methods are often limited by restrictive assumptions \cite{Hora2024}. LSTM networks have also shown good reconstruction performance with sparse turbulence data \cite{Poulinakis2024}. 

A particular application for which data-driven methods have yielded recent progress is the detailed reconstruction of turbulent flows near interfaces, where measurements at the interface give input to the learning algorithm. Such cases include reconstructing the bottom topography \cite{gakhar2022} and full subsurface velocity fields \cite{xuan2023} from free surface measurements, and reconstructing wall-bounded turbulence from wall measurements only. The latter use CNNs for large-scale reconstruction \cite{Guastoni2021,guemes2019}, CNN-based variational autoencoders \cite{Hora2024}, or GANs \cite{Guastoni2021, Cuellar2024}, yielding great reconstruction performance compared to linear methods. 
For the reconstruction of subsurface flow fields from free surface data, Xuan \& Shen (2023) \cite{xuan2023} applied a CNN on data from DNS, using surface elevation and the surface velocity field as input. The results are promising and outperform previous linear reconstruction models, indicating that turbulent flow fields in, say, a river or lake, may indeed be inferred from the water surface only. In this case, however, the CNN is applied image-by-image for reconstruction, hence independent of temporal dynamics. Moreover, it is pointed out that CNN methods may underestimate the fluctuating amplitudes of large-scale structures. Beyond these issues, CNN-based approaches often prove to be computationally expensive and either require dense surface fields or struggle with noisy experimental data \cite{discetti2022}, leaving a gap for a lightweight model that can operate on sparse field data and generalize across simulation and laboratory flows. 
A more general issue with ``sensing" turbulent flow fields from interface measurements is the fall in accuracy with increased distance from the measured interface, due to less correlation of the flow and interface far from the latter. 
Hence, closing the gaps of both efficiency and accuracy is essential to take subsurface flow sensing
to a level of practical utility, i.e., for practical remote sensing of river turbulence and subsequent gas‑exchange estimation.

\begin{figure}
    \centering
    \begin{overpic}[width=\textwidth, trim={0cm 0cm 2.2cm 0cm}, clip, grid=off, unit=2bp,tics=2]{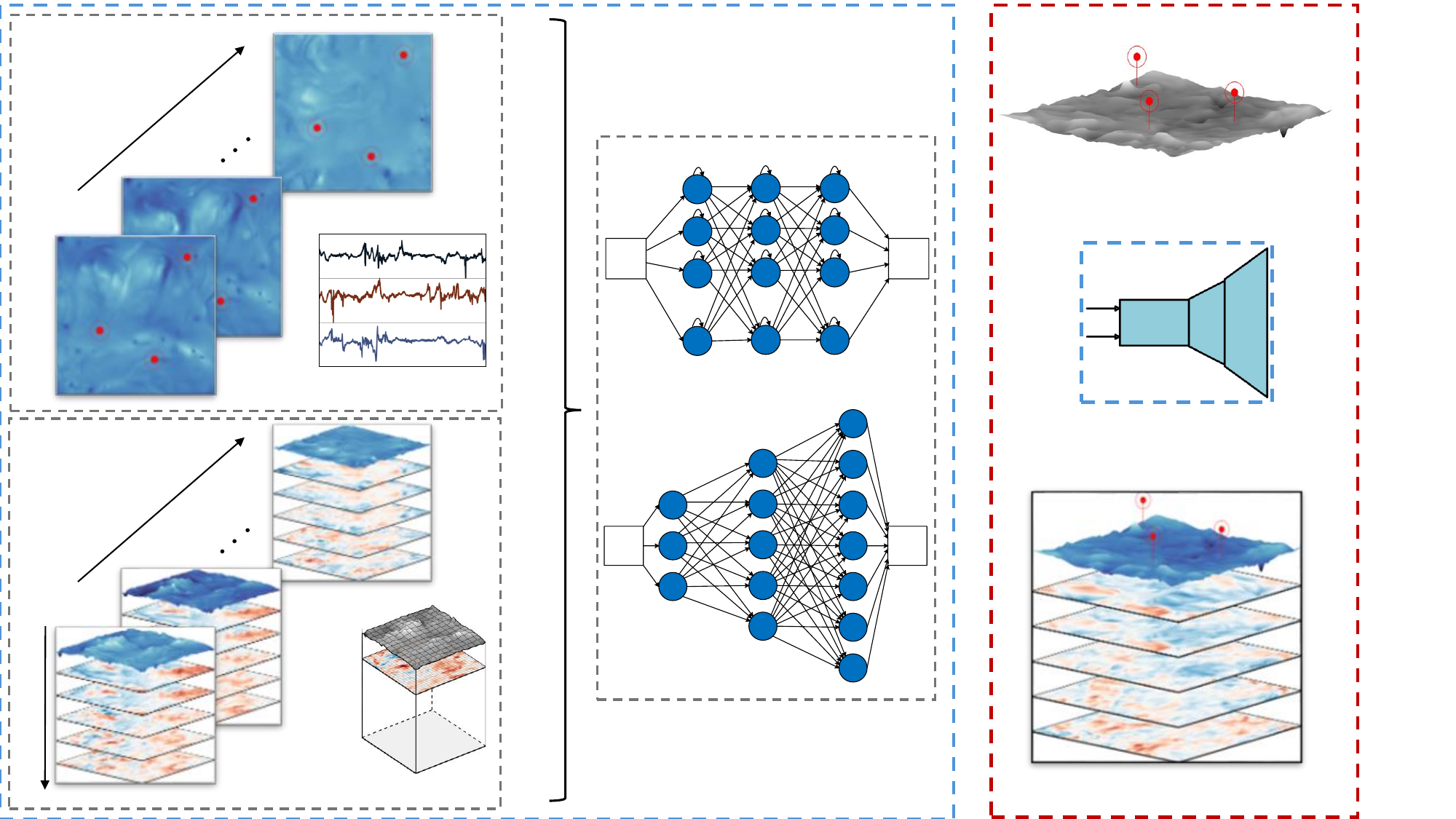}
    \put (9,51) {$t$}
    \put (9,22) {$t$}
    \put (18,31) {Sensor readings}
    \put (18, 1.5)  {Full data arrays}
    \put (56, 57) {TRAINING}
    \put (74, 57) {DEPLOYMENT}
    
    \put (45,31)  {LSTM}
    \put (45,10)  {SDN}
    \put (1,4) {\rotatebox{90}{Depth}}
    \put (56,30) {\rotatebox{90}{\Large$\leftarrow$}}
    \put (86,44) {\rotatebox{90}{\Large$\leftarrow$}}
    \put (86,25) {\rotatebox{90}{\Large$\leftarrow$}}
    \put (44.7,40.4) {$W$}
    \put (65.7,40.4) {$\hat{W}$}
    \put (44.7,19.3) {$\hat{W}$}
    \put (65.7,19.3) {$V$}
    \put (90, 46) {Input}
    \put (90, 27.5) {Model}
    \put (90, 1.5)  {Output}
    \put (78,26) {{$US\tilde{V}^T$}}
    \put (38,10) {\rotatebox{90}{$V$-matrix from SVD + sensor data $W$}}
    \end{overpic}
    \caption{Illustration of the SHRED architecture, training and deployment phase. Left: 
    \ajo{
    Input and output to SHRED, consisting of
    }%
    time series from three sensors 
    \rjo{and}{
    ($W$) combined with
    }%
    the time dynamics of the DNS or 
    \rjo{PIV}{
    PIV/profilometry
    }%
    data (%
    \ajo{%
    compressed %
    }%
    $V$-matrix from SVD of slice planes)
    \rjo{%
    to form matrix $W$.
    }{, respectively.}
    Middle: 
    \djo{Matrix }
    $W$ processed by long-short time memory%
    \djo{ and shallow decoder network}\ajo{%
    ; the resulting latent representation $\hat{W}$ mapped onto $V$ by the shallow decoder network}. Right: The model is deployed by taking new measurements of the surface%
    \ajo{%
    , outputting the reconstructed time dynamics $\tilde{V}$%
    }%
    . After processing 
    \djo{we reconstruct }
    the full surface and velocity slice planes
    \ajo{
    are reconstructed by matrix multiplication of $\tilde{V}^T$ with the compressed SVD-matrices $U$ and $S$.
    }%
    }
    \label{fig:SHREDarch}
\end{figure}

A recently developed method\ajo{ by Williams et al.~\citep{williams2024}}, the SHallow REcurrent Decoder (SHRED), takes inspiration from recent data-driven networks to reconstruct spatiotemporal fields from sparse sensor measurements of a single quantity\djo{\cite{williams2024}}. It is based on the principle of separation of variables, where the temporal dynamics is learned separately with an LSTM encoder, while spatial structures are recovered with a shallow decoder network (SDN), see illustration in Fig. \ref{fig:SHREDarch}. More specifically, time series of sparse sensor measurements of a single quantity field (e.g. pressure, surface elevation, velocity) are given as input to the LSTM. It then constructs a latent space representation of the time dynamics of the field that the SDN is trained to map onto all correlated fields of interest, or their compressed representations~\cite{Faraji2025,tomasetto2025reducedPub}. This is possible because of Takens' embedding theorem, which states that as long as each time series is treated as a delay coordinate embedding, the underlying flow dynamics attractor is contained in the embedding \cite{Takens1981}. Hence, this enables the decoder to learn a smooth mapping to the full spatial field. The compressive training is critically enabling as it allows for rapid and robust training on laptop-level computing for a variety of high-dimensional, multi-physics systems such as plasma Hall thrusters
~\cite{Faraji2025}, nuclear reactors \cite{riva2025}, circulating fuel reactors \cite{introini2025modelsexperimentsshallowrecurrent}, and reduced order models~\cite{tomasetto2025reducedPub}. For the first two applications, the models featured 14 and 21 coupled PDEs, respectively, with only a single field measured with three randomly placed sensors. SHRED can also be used with mobile sensors~\cite{Ebers2024}, and can even be combined with the {\em sparse identification of nonlinear dynamics} (SINDy) or Koopman methods for model identification from sensing alone \cite{gao2026}. The ability to infer and sense fields distinct from the one being measured makes SHRED a suitable candidate in the quest for remote sensing of subsurface flows, as in rivers and oceans. Hence, SHRED is a promising new addition to the landscape of data-driven methods for fluid flow sensing. It is therefore of great interest to apply SHRED in the context of free-surface turbulence and subsurface reconstruction.

This paper presents a case study of SHRED deployed to flow data from turbulent free-surface flow,
\replace{obtained}{
made available to us
}%
from two separate sources: direct numerical simulations
\add{
performed at the University of Minnesota, previously published in \cite{aarnes2025,aarnes2025b}
}%
and experiments
\add{from a recent laboratory study at the Norwegian University of Science and Technology \cite{babiker2026,babiker2025}.}
In both 
\replace{cases, }{
datasets}%
\akr{, }isotropic, homogeneous turbulence 
\replace{is }{was} 
created well below the surface whence it naturally diffuses towards the surface.  Fig.~\ref{fig:SHREDarch} exhibits the SHRED algorithm in relation to our mapping of surface measurements to subsurface flows.
\replace{The present flow case }{
This category of free-surface turbulent flow
}%
is well documented in the turbulence literature, both experimentally \citep{brumley1987,Chiapponi2012, herlina_experiments_2008,Hopfinger1976,Jamin2024,Lacassagne2017, McKenna2004, Thompson1975, Variano2013} and numerically \citep{guo2010,herlina2014, herlina2019, babiker2023,aarnes2025}, and can be considered fundamental to our understanding of free-surface turbulence dynamics. The relevance to a number of naturally occurring flows is immediate, such as the near-surface flow of a river or the upper layer of the ocean when the free-surface shear stress due to wind is negligible.
We present the performance of SHRED on 
\add{
such free-surface turbulent flow
}%
data 
\delete{from this flow case }
as a proof of concept of data-driven sensing of subsurface flows from surface measurements only, in a highly challenging turbulent environment characterized by large intermittent structures and absence of periodic features. 
In stress testing SHRED for different Reynolds numbers from DNS and laboratory experiments, we aim at connecting the SHRED performance to real non-ideal flows as typically found in 
\replace{real-life cases of}{
natural environments such as
}%
rivers and oceans. 
Moreover, our results highlight the significant experimental gain of reconstruction of subsurface flows from surface-only measurements, as a step towards more accurate non-intrusive measurement techniques.

The paper is organized as follows: In \S\,\ref{sec:shred} we \rjo{introduce}{describe} the reconstruction model SHRED and in \S\,\ref{sec:data_generation} we 
\replace{provide details on }{
summarize
}%
the simulations and experiments from which our training data and reconstruction datasets 
\replace{are gathered, as well as}{originate,}
as well as data compression and performance metrics used. We present our results in \S\,\ref{sec:results}, among them a set of error metrics which discloses how the performance of SHRED varies with distance from the reconstructed plane to the free surface. Lastly, we draw conclusions in \S\,\ref{sec:conclusion} and include an appendix with details on parameter tuning thereafter.

\section{SHRED: SHallow REcurrent Decoder}\label{sec:shred}

SHRED \cite{williams2024} is a generalization of the separation of variables methods for solving partial differential equations (PDEs)~\cite{folland2020introduction}. Separation of variables is also the underlying technique used for many spectral integration methods for solving PDEs~\cite{kutz2013book}. The method assumes that a solution can be separated into a product of time and space functions $u(x,t)= T(t) X(x)$.  The solution reduces the PDE into an ordinary differential equation for time $T(t)$  and a boundary value problem for space $X(x)$.

To demonstrate the method, consider the constant coefficient linear PDE 
\begin{equation}
 \dot{u} = {\cal L} (\partial_x, \partial^2_x, \cdots ) {u}
 {\label{eq:linearPDE}}
\end{equation}
where $u(x,t)$ specifies the spatiotemporal field of interest subject to the physics imposed by the operator  ${\cal L}$. Sample initial conditions (IC) and boundary conditions (BCs) are given by
\begin{subequations}
\begin{align}
\text{IC:}&&\quad  u(x,0)=u_0(x) \\
\text{BCs:}&&\quad  \alpha_1 u (0,t) +\beta_1 \partial_x u(0,t) = g_1(t) \,\, \text{and} \nonumber \\  && \alpha_2 u(L,t) + \beta_2 \partial_x u(L,t) = g_2(t).
\end{align}
\label{eq:ICBC}
\end{subequations}
This may be generalized to systems of several spatial variables, a system with no time dependence, or a coupled system of equations.  The linear operator ${\cal L}$ specifies the spatial derivatives, which in turn model the underlying physics of the system.  Simple examples of ${\cal L}$ include ${\cal L}= c \partial_x$ (the one-way wave equation) and ${\cal L}= \kappa \partial^2_x$ (the heat equation)~\cite{kutz2013book}.  

The earliest solutions of linear PDEs assumed separation of variables whereby $u(x,t)=\exp(\lambda t) X(x)$ was a product of a temporal (exponential) function multiplied by a spatial function.  The parameter $\lambda$ is in general complex and specifies the eigenfunction solution
\begin{equation}
    u(x,t) = \sum_{n=1}^{N} a_n \exp(\lambda_n t) \phi_n(x)
    \label{eq:ef}
\end{equation}
where $\phi_n(x)$ are the eigenfunctions of the linear operator and $\lambda_n$ are its eigenvalues (${\cal L}\phi_n(x) = \lambda_n \phi_n(x)$).  Here a finite dimensional approximation $N$ is assumed, which is standard in practice for numerical evaluation. The solution of Eq.~(\ref{eq:ef}) is a general solution which models all possible solutions.  To specify a unique solution, initial conditions $u(x,0)=u_0(x)$ are typically imposed in order to uniquely determine the coefficients $a_n$.  Specifically, at time $t=0$, Eq.~(\ref{eq:ef}) becomes
\begin{equation}
    u_0(x) = \sum_{n=1}^{N} a_n  \phi_n(x) .
\end{equation}
Taking the inner product of both sides with respect to $\phi_m(x)$ and making use of orthogonality gives
\begin{equation}
   a_n =\langle u_0(x), \phi_n(x) \rangle .
\end{equation}

As an alternative to specifying the initial data at all spatial points, SHRED instead specifies measurements at a single spatial (sensor) location $x_s$, but with a temporal history.  Multiple point measurements can be used as well without loss of generality.  Thus if SHRED has $N$ temporal trajectory points, this gives at each time point of the measurement a constraint:
\begin{equation}
    u(x_s, t_j) = \sum_{n=1}^{N} a_n 
    \exp(\lambda_n t_j) \phi_n(x_s) 
    \,\,\,\,\,\, \mbox{for} \,\,\, j=1,2,\cdots N .
\end{equation}
This results in $N$ equations for the $N$ unknowns $a_n$. Specifically, the $N\times N$ system of equations ${\bf A} {\bf x} = {\bf b}$ is prescribed by the vector components $x_k=a_k$ and $b_k = u(x_s,t_k)$ and matrix components $(a_{kj}) = \exp(\lambda_k t_j) \phi_k(x_s)$.  As with the initial condition (\ref{eq:ICBC}a), the time trajectory of measurements at a single location uniquely prescribes the solution. This analysis can easily be generalized to include multiple sensor measurements at a single time point. Thus if there are two measurements at a given time $t_j$, then only $N/2$ trajectory points are needed to uniquely determine the solution.  Likewise, three sensor measurements at a given time require $N/3$ trajectory points.
In addition to stationary sensors measurements, one can also consider mobile sensors whereby the measurement of the system is a different locations over time~\cite{Ebers2024}:  $x_s = x_{s(t_j)}$. The above arguments are easily modified so that the vector component $b_k = u(x_{s(t_j)},t_k)$ and matrix components $(a_{kj}) = \exp(\lambda_k t_j) \phi_k(x_{s(t_j)})$.

Thus temporal trajectory information at a single spatial location, or with a moving sensor, is equivalent to knowing the entire initial condition.  SHRED provides a generalization to separation of variables $u(x,t) = T(t)X(x)$ by encoding time with a time sequence model such as an LSTM model and a decoder model for full-state reconstruction of space.  Rigorous theoretical bounds of SHRED are difficult to achieve, much like analytic and numerical solutions are difficult to rigorously bound in computational PDE settings. But in the linear limit, the above arguments show explicitly why SHRED is guaranteed to work and recover the full spatiotemporal field exactly.

\subsection{Nonlinear PDEs}
 
For nonlinear PDEs of the form
\begin{equation}
 \dot{u} = N (u, \partial_x u, \partial_{x}^2 u, \cdots ), 
 {\label{eq:nonlinearPDE}}
\end{equation}
numerical methods are commonly used to generate solutions subject to the initial and boundary conditions~(\ref{eq:ICBC}).  Consider a spectral solution technique~\cite{kutz2013book} whereby numerical solutions are approximated by a spectral basis
\begin{equation}
    u(x,t) = \sum_{n=1}^{N} a_n (t) \phi_n(x) .
    \label{eq:spectral}
\end{equation}
Typical examples of spectral techniques include using Fourier modes or Chebychev polynomial for $\phi_n(x)$.  This spectral decomposition turns the PDE into a systems of $N$ coupled ordinary differential equations for $a_n(t)$:
\begin{equation}
  \frac{da_n}{dt}=f_n (a_1, a_2, \cdots, a_N) \,\,\,\,\,\, \mbox{for} \,\,\, n=1, 2, \cdots N 
  \label{eq:an}
\end{equation}
The solution of the $N$-dimensional differential equation has $N$ unknown constants of integration that are typically uniquely determined by applying initial conditions and orthogonality in Eq. (\ref{eq:spectral})
\begin{equation}
    a_n(0)= \langle u_0(x), \phi_n(x) \rangle   . 
\end{equation}
As with the separation of variables solution, we can instead assume that we can construct a general solution  for Eq. 
(\ref{eq:an}) which has $N$ constants of integration. The constants of integration can be determined by requiring the solution to satisfy $N$ temporal trajectory points, giving at each time point of the measurement:
\begin{equation}
    u(x_s,t_j)=\sum_{n=1}^N a_n(t_j)\phi_n(x_s)  \,\,\,\,\,\, \mbox{for} \,\,\, j=1,2,\cdots N .
\end{equation}
This gives $N$ constraints for the $N$ unknown constants of integration, thus uniquely determining the evolution of the $a_n$ in Eq. (\ref{eq:an}).  Mobile sensors can also be used to enforce the constraints required for a unique solution. \add{For details on how this extends to coupled PDEs, see \cite{williams2024}.}



%
%
\delete{where $u(x,t)$ and $v(x,t)$ specifies the spatiotemporal fields of interest. The PDEs can instead be written in the form}
%
\delete{where Eq. (\ref{eq:linearPDE2}a) is differentiated with respect to time and Eq. (\ref{eq:linearPDE2}b) is used in order to write the PDEs as a function of $u(x,t)$ alone.  Thus, knowledge of the field $u(x,t)$ alone is capable of constructing the solution fields $u(x,t)$ and $v(x,t)$.  For this second-order (in time) PDE, both an initial condition $u(x,0)$ and an initial {\em velocity} $\dot{u}(x,0)$ require specification in order to uniquely determine the solution.  As with the previous arguments, a time trajectory embedding of $2N$ measurements can be used to uniquely determine the solution.}


\subsection{\ajo{Training and deployment}}

We \ajo{train and }deploy SHRED on 
\ajo{
the four
}%
datasets 
\ajo{
detailed in \S\ \ref{sec:data_generation}, each
}%
consisting of simultaneous velocities and free-surface elevation resolved in space and time, 
\delete{from the two DNS cases and two experimental cases}%
according to the principles laid out in Fig.\ \ref{fig:SHREDarch}. We input time series of surface elevation from 
\djo{three }%
randomly placed surface sensor points into a two-layer LSTM. 
\ajo{
The input matrix $\bW$ is an $n_s\times n_t$ matrix, where $n_s$ is the number of sensors and $n_t$ is the length of the time series. For all results we report, $n_s = 3$, excepting the discussion of influence of the number of sensors on reconstruction accuracy in Appendix \ref{sec:app2}. The time series length $n_t$ depends on dataset and whether we are in the training or deployment stage. %
}%
The LSTM encodes these input sequences 
\ajo{
from the surface sensors
}%
into a latent representation of their temporal dynamics%
\ajo{%
, denoted $\hat{\bW}$ in Fig.\ \ref{fig:SHREDarch}, of size $n_{\text{lat}}\times n_t$, with $n_{\text{lat}}=64$ latent state dimensions%
}%
. This latent vector is then passed to a shallow decoder network (SDN), which maps it onto the
\ajo{%
compressed $\bV$ matrices, obtained from SVD and rank truncation, for the surface elevation and the subsurface velocity fields
}%
across depth. 
\dkr{We do this in compressed space, by feeding into the SDN the compressed $\bV$ matrices of the SVD decomposition for the surface elevation and the subsurface velocity fields.}
These fields are used in training and validation not to learn the subsurface time dynamics but only to learn the mapping of the surface time dynamics onto the subsurface fields. This is an essential detail in the context of remote sensing. To output the full spatiotemporal fields, we save the compressed $\bU$ and $\bS$ matrices from the \akr{SVD of the} training data, and matrix multiplication with 
\rjo{$\bV^*$
}{
the reconstructed time dynamics, $\tilde{\bV}$,}
then yields the reconstruction fields. This is computationally cheaper than feeding in the full spatiotemporal fields, although that is also possible. 

Each SHRED model is independently trained on one continuous dataset at a time, that is, no cross-case or cross-domain training is used. For each 
\replace{case, }{
dataset,
}%
we randomly split the data into 80\% training, 10\% validation, and 10\% testing snapshots, all within the same time series. An ensemble of SHRED models is trained in each case to assess convergence and uncertainty for each run. \ajo{Note that using SHRED as a forecasting tool, where the training and test data are separated in a temporal sequential manner, is beyond the scope of the present study. For an example of SHRED used for forecasting, see  \cite{kutz2024shallow}.
}

SHRED uses a two-layer LSTM encoder and a two-layer SDN decoder (with no dropout). We use the Adam optimizer with an initial learning rate of $10^{-3}$, and train on mini-batches of 64 time snapshots. The loss function is the mean squared error (MSE), computed between reconstructed and true \rkr{velocity fields}{$\bV$ matrices} in the compressed domain. Details can be found in \footnote{\href{https://github.com/krissmoe/SHRED-turbulence-sensing}{https://github.com/krissmoe/SHRED-turbulence-sensing}} the \href{https://github.com/krissmoe/SHRED-turbulence-sensing}{GitHub} code repository. 

\begin{figure}
    \centering
    \begin{subfigure}[b]{0.47\textwidth}
        \includegraphics[width=\linewidth]{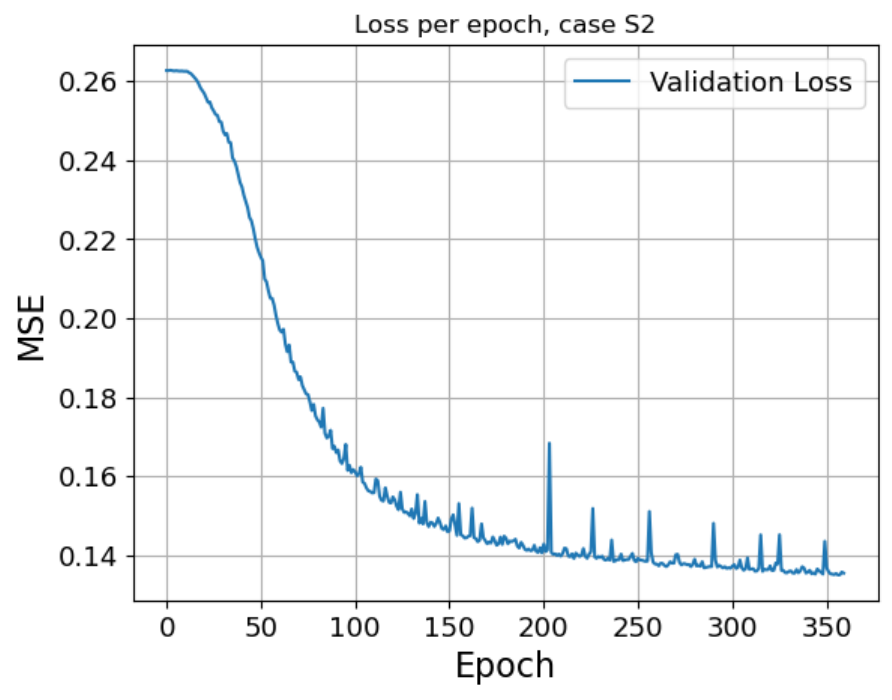}
        
    \end{subfigure}
    \hfill
    \begin{subfigure}[b]{0.47\textwidth}
        \includegraphics[width=\linewidth]{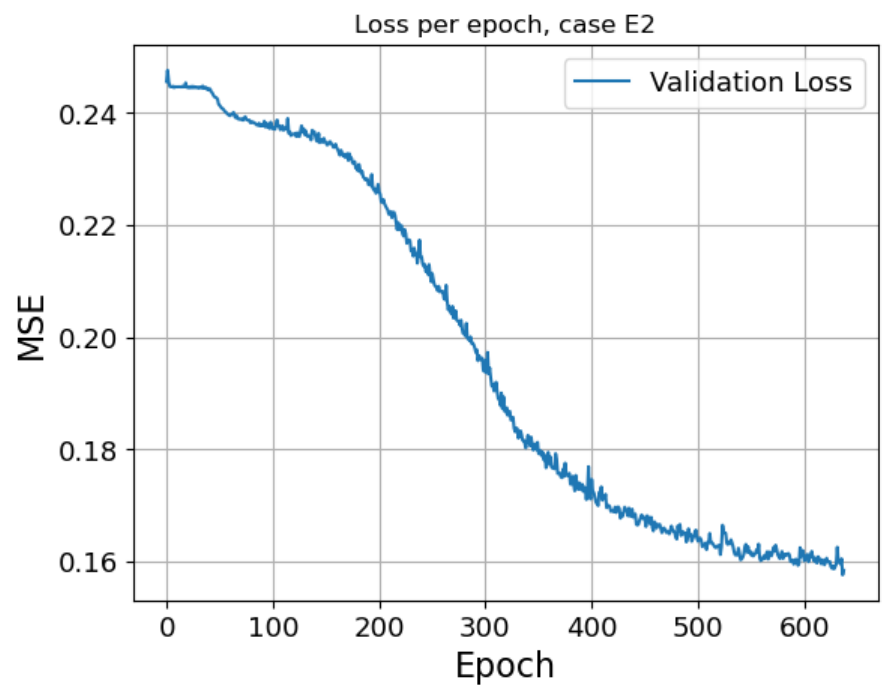}
        
    \end{subfigure}
    \caption{Example of typical MSE loss profile for the validation dataset 
    \delete{from case}%
    S2 (left) and E2 (right).}
    \label{fig:loss_landscape}
\end{figure}

Fig.\ \ref{fig:loss_landscape} shows a typical validation MSE loss curve for 
\replace{cases}{
datasets
}%
S2 and E2\ajo{ (flow metrics in Table \ref{tab:case_details})}. We observe a steep drop in validation after around $40$ epochs for case S2, and around $200$ epochs for 
\delete{case} 
E2, and a convergence typically occurs after a few hundred epochs. Running on GPU, a full SHRED run 
\ajo{
(excepting the data compression step)
}%
of surface paired with a single velocity plane, takes 1-2 minutes on a regular desktop computer. A simultaneous run with all planes (only possible with the datasets 
\delete{in cases} 
S1 and S2) takes up to 10 minutes.

\section{Data generation, collection, and evaluation}
\label{sec:data_generation}
\subsection{DNS and experiments}
We deploy SHRED to four different datasets, two sets of data from direct numerical simulations and two from experimental data which utilize a combination of PIV and profilometry to capture both the surface displacement and the subsurface dynamics. 
The data have been thoroughly documented 
elsewhere and only a brief outline is repeated here. Details on the simulation parameters, numerical schemes, and grid resolution criteria based on grid independence studies can be found in \cite{guoGenerationMaintenanceWaves2009,guo2010} and \cite{xuan2019,xuan2022}, with specifics related to the dataset in use given in \cite{aarnes2025}. Details of the data capture method in the experiment and the experimental setup can be found in \cite{ali2025forthcomming} and
\cite{babiker2025}, respectively.

\begin{figure}
    \subfloat[Simulation]{
        \begin{overpic}[width=0.44\textwidth, trim={2.7cm 0.5cm 0.5cm 0.5cm}, clip, grid=off]{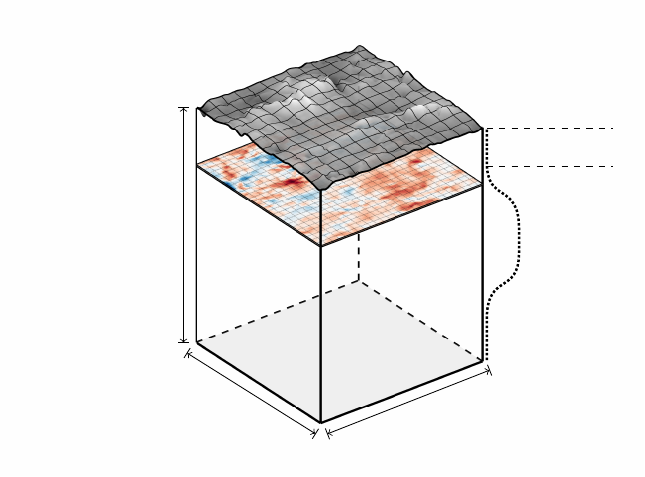}
            \put (0,48) {$H$}
            \put (16,8) {$L_x$}
            \put (58,5) {$L_y$}
            \put (81,44) {$f(z)$}
            \put (73,64) {Free region}
        \end{overpic}
    } 
    \subfloat[Experiment]{
        \includegraphics[width = 0.35\textwidth]{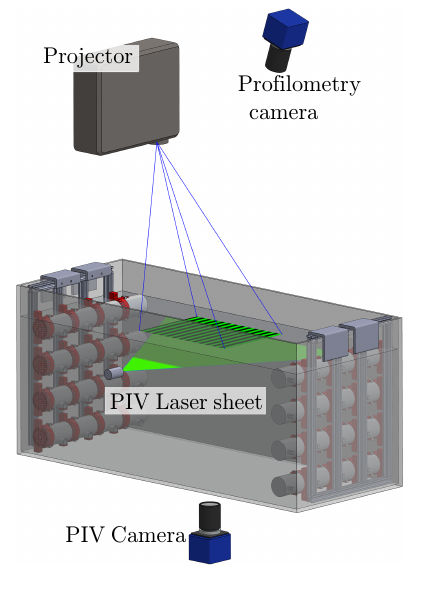}
    }
    \caption{(a): Schematics for the simulation domain, with undulating grid following the free surface, depth-dependent forcing function $f(z)$ ($=0$ in the free region), and periodic boundaries along the horizontal directions. 
    \add{
    Adapted from Fig.\ 2 of \cite{aarnes2025}.
    }%
    (b): 
    \replace{Experimental setup, where turbulence is}{
    Facility used in experiments, reprinted from Ref.~\cite{babiker2025} with permission. Turbulence was
    }%
    generated by randomly actuated jets and data is captured by profilometry (projector and camera 1) and PIV (sheet-making optics and camera 2; laser and mirrors not in the frame). }
    \label{fig:DNSexp}
\end{figure}

The simulations mimic experiments of free-surface homogeneous turbulence where turbulence is generated far beneath the water surface and diffuses towards the surface through self-interaction of turbulence vortices and viscous diffusion, with turbulence generated by an oscillating grid \citep[see, e.g.,][]{Thompson1975, brumley1987, herlina_experiments_2008,Lacassagne2017} or jets with zero net flow \citep[e.g.][]{Variano2013,Jamin2024, ruth2024, babiker2025}.
In the simulations \cite{aarnes2025,xuan2019, guo2010}, the non-dimensional, incompressible Navier--Stokes equations and continuity equation 
\replace{are}{
were
}%
solved on a three-dimensional domain, with periodic boundary conditions in horizontal directions, a free-slip boundary at the bottom, and a free surface boundary on the top. The latter is enforced by the dynamic and kinematic boundary conditions and tracked by a surface-adhering grid that undulates in the vertical direction as the surface elevation varies. Turbulence is generated in the center region of the domain by random linear forcing \cite{lundgren2003, rosalesLinearForcingNumerical2005}, modulated by a depth-dependent function which is strongest in the center of the domain and drops to zero well below the surface-influenced region [see Fig. \ref{fig:DNSexp}(a)]. The top region, denoted the `free region', is free from random forcing. Only data from the free region are used in the analysis. Surface tension is neglected in the simulations, which has little effect on the degree of correlation between surface and bulk as long as the flow is in the high-Weber-number regime (see extensive discussion in \cite{aarnes2025}). \ajo{The mesh sizes of the two DNS datasets are \(128 \times 128 \times 348\) and \(256 \times 256 \times 660\), for the low and moderate Reynolds number case, denoted S1 and S2 in Table \ref{tab:case_details}, respectively.}

\begin{table}
    \centering
\add{
     \begin{tabular}{p{4em} p{3em} p{3em} p{3em} p{3em} p{3em} p{3em} p{3em} p{3em} p{2em}}
    \hline
        Dataset   &  $\mathrm{Re}_\infty$ & $\mathrm{Fr}_\infty$ & $\mathrm{We}_\infty$ & $\mathrm{Re}_\lambda$ &
      $L_\infty$  &  $\lambda_T$ &$L_K$ & $\urms$ \\
      
        &&&&&[mm]&[mm]&[mm]&[mm/s]\\
        \hline
        S1 & {267} & {0.009} & $\infty$ & {47} & {21.7} & {7.59} & {0.56} & {6.15}  \\     
        S2 & {782} & {0.010} & $\infty$ & {84} & {43.0} & {9.18} & {0.51} & {9.10}  \\

        E1 & {7636} & {0.040} & {5.2}     & {279} & {77.1} & {5.5} & {0.17} & {49.5}  \\

        E2 & {17369} & {0.050} & {18.1}   & {397} & {115.4} & {5.3} & {0.13} & {75.3} \\
        \hline
    \end{tabular}
}
     \caption{Flow and turbulence properties for 
     \add{
     data from 
     }%
     simulations (S) and experiments (E). From left: Case label,
     turbulent Reynolds number, turbulent Froude number ($ \mathrm{Fr}_\infty = u'/ \sqrt{2 g L_\infty}$), turbulent Weber number ($\mathrm{We}_\infty = \rho u^{\prime 2}L_\infty / \sigma$), Taylor Reynolds number, integral length scale, Taylor length scale, Kolmogorov length scale. 
     \replace{Length scales from simulations are normalized with the reference length scale used in the DNS simulations, as detailed in \cite{aarnes2025}. Length scales and velocity values from experiments are given in SI units.}{
     Details on the calculation of the tabulated values can be found in \cite{babiker2025}.
     }
     }
    \label{tab:case_details}
\end{table}

\replace{In the experiments, turbulence is }{
The experimental flow data are from the experiment of Babiker \emph{et al.}~\cite{babiker2025} where full details can be found. Turbulence was
}%
generated by randomly actuated water jets below the water surface in a rectangular tank (see Fig.~\ref{fig:DNSexp}(b); note that only the two bottom rows of the jets 
\replace{are}{
were
}%
active).
\delete{, see [4].}
With a 
\delete{novel}
data collection technique 
\add{
developed recently by Semati \emph{et al} \cite{ali2025forthcomming}
}%
that combines surface profilometry with particle image velocimetry (PIV), the surface displacement and velocity in a subsurface horizontal plane 
\replace{are }{
were
}%
captured simultaneously.
\delete{(see \cite{ali2025forthcomming}; for alternative methods of measuring the surface and the subsurface simultaneously, see 
[refs].}%
Unlike in the simulations, only a single velocity plane 
\replace{is}{
was
}%
resolved at 
\replace{each time step. The laser sheet in use is varied between}{
a time, serially measuring at
}%
four different 
\add{
depth
}%
levels between ensembles to get estimates for depth dependency. During characterization of the experiment, a vertical laser sheet 
\replace{is }{
was
}%
used and, hence, well resolved root-mean-square measures of the velocity are obtained. The 
\replace{experiments measured surface profilometry with each corresponding}{
experimental dataset consists of the measured free-surface elevation directly above each measured 
}%
PIV velocity field for 20 time intervals, or ensemble cases, each with a 1-minute duration. 
\replace{In order for the cases to be independent, there was a minute pause in between ensemble cases. Due to noise and issues in the profilometry for a few cases, only }{
Of these, 
}%
$14$ 
\replace{ensemble cases are }{
were
}%
considered 
\replace{usable }{
suitable
}%
data for the SHRED analysis
\add{
we perform here.
} 

\replace{Case details}{
Flow metrics
}%
for the four datasets are listed in Table \ref{tab:case_details}. There, two additional dimensionless numbers are included alongside the Reynolds number. These are the 
\ajo{
Froude
}%
number and the Weber number. While the Reynolds number expresses the ratio of inertial to viscous forces, the Froude number takes into account the gravitational acceleration, $g$, and the Weber number expresses the influence of inertial forces to forces due to surface tension, $\sigma$.
    \delete{ Take note that directly comparing DNS data and data from experiments is not trivial (see extensive discussion in \cite{babiker2025}). With non-dimensional DNS data the closest we get to a direct comparison is the turbulent Reynolds number and the Taylor Reynolds number. Yet even here, alternatives in how to compute length scales introduce discrepancies. Consider, for example, how computing integral length scales from dissipation and characteristic velocity (used in the DNS data) or by structure functions (used in the experimental data) introduce a difference in the length scale, and thus also in the Reynolds number of turbulence, $Re = 2\tilde{u}L_\infty/\nu$, where $\tilde{u}$ is the representative velocity (i.e., the root-mean-square velocity, see \cite{tennekes1972first}), $L_\infty$ is the integral scale and $\nu$ is the kinematic viscosity.} 
Details on how these quantities are calculated in practice can be found in 
\djo{refs.\ }%
\cite{aarnes2025}. 
\ajo{
Extensive details on the comparison of these DNS and experimental datasets by dimensionalization of the DNS data and appropriate estimation of integral scales and reference depths can be found in \cite{babiker2025}.
}

For brevity, we limit our scope to reconstructing one component of the velocity field, a horizontal velocity component denoted $u$. Since the turbulence is horizontally isotropic, the choice of horizontal axis is without consequence. 
In our analysis, we refer to the root-mean-square velocity as a representative quantity, which we define as

\begin{equation}\label{eq:urms}
    \urms = \left(\frac1{mn}\sum_{i=1}^m \sum_{j=1}^n u_{x_i,y_j}^2 \right)^\frac12 \, ,
\end{equation}
where the indices 
\ajo{
$x_i$ and $y_j$
}%
denote the positions in the horizontal plane, referring to the discrete data matrices in the DNS and PIV data grids, respectively, of dimension $m\times n$.
Thus, $\urms$ varies as a function of depth, $z$, and time, $t$. Since we always consider only the single velocity component $u$, $\urms^2$ should not be confused with twice the kinetic energy density. 

\subsection{Compression}
\label{sec:compression}

When reconstructing subsurface \ajo{turbulent }flows, data compression 
\rjo{
is necessary for
}{may facilitate} 
good reconstruction, 
\rjo{
but
}{and} 
is also justified by the nature of turbulence. We use singular value decomposition (SVD) to evaluate the relevant modes and scales for the compression of the high-dimensional data to a low-rank representation capturing the essential dynamics.%
\ajo{
The SVD is performed on each plane and flow variable separately, ensuring no data leakage across the depth of the flow from the decomposition.
}

The DNS and experimental datasets contain turbulence data with a wide range of spatial scales. From turbulence theory, it is well known that energy is injected at the largest scales, before cascading to smaller and smaller scales in the inertial $k^{-5/3}$ range, and dissipating to heat at the Kolmogorov length scale. The majority of the turbulent kinetic energy is carried by the largest structures, which dominate the transport of heat, momentum, and mass in the flow. When we use SHRED to reconstruct the turbulent free-surface flow from surface measurements, we aim to accurately capture the large scales and avoid overfitting to the intermittent and unpredictable small scales of the turbulent spectrum. We achieve this by compressing the data by 
\djo{singular value decomposition}%
SVD, keeping only a small amount of the decomposed data. In addition to increased accuracy, using SHRED with rank-reduced data have the advantage of a very significant speed-up of training, validation, and reconstruction. (For details on the SVD algorithm, see, e.g., \cite{BruntonKutz2019, StrangGilbert2019Laal}.)

\begin{figure}[!htbp]
    \centering
    \begin{tikzpicture}
        \node[anchor=south west, inner sep=0] (image) at (0,0) {
            \includegraphics[width=\textwidth, trim={0.4cm 0cm 0cm 0cm}, clip]{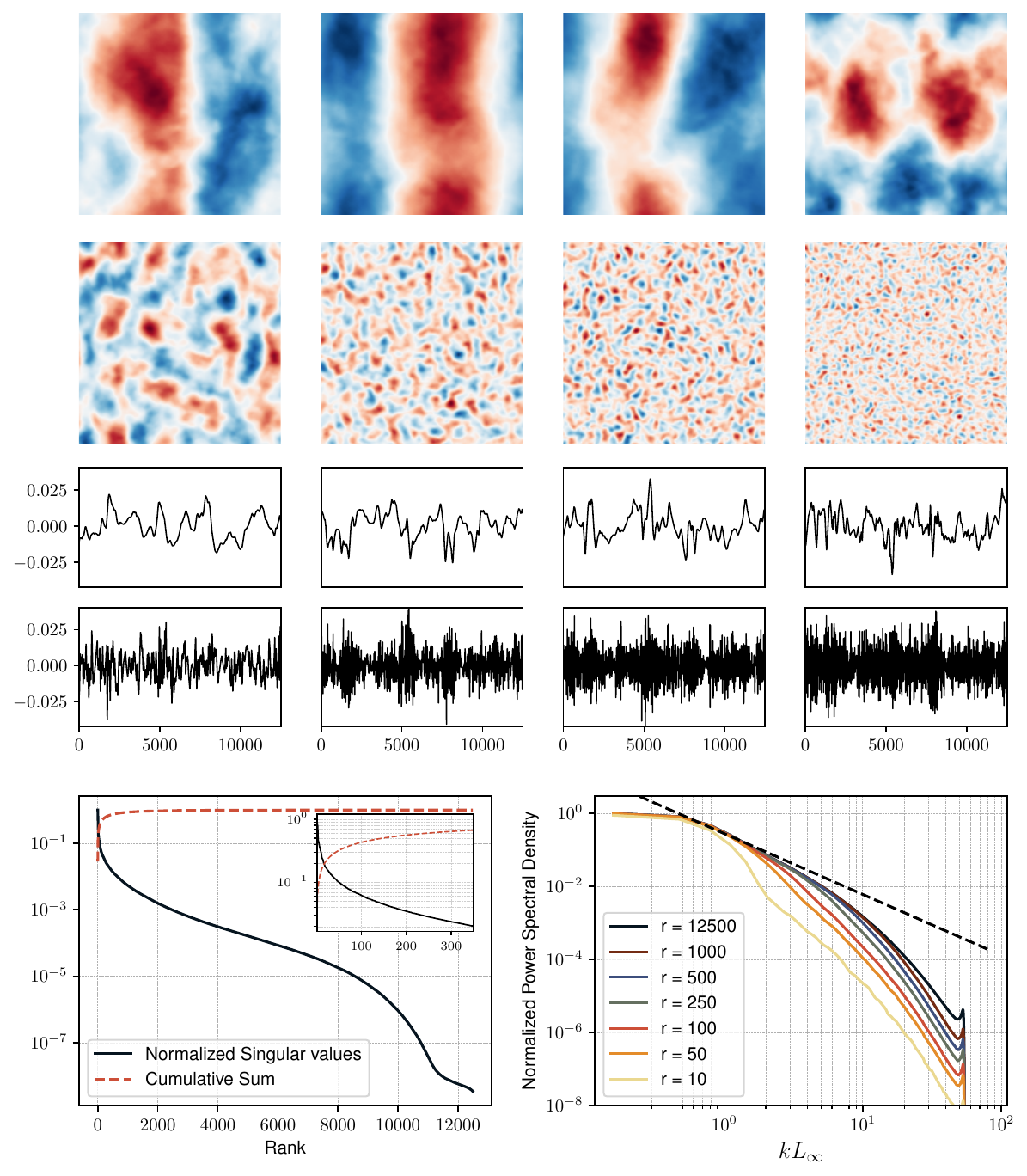}
        };

        \begin{scope}[shift={(image.south west)}, x={(image.south east)}, y={(image.north west)}]
            
            
            \labelbox{}{0.10, 0.96}{$u_1$}
            \labelbox{}{0.34, 0.96}{$u_2$}
            \labelbox{}{0.585, 0.96}{$u_3$}
            \labelbox{}{0.825, 0.96}{$u_{10}$}
            \labelbox{}{0.10, 0.765}{$u_{50}$}
            \labelbox{}{0.34, 0.765}{$u_{250}$}
            \labelbox{}{0.59, 0.765}{$u_{500}$}
            \labelbox{}{0.835, 0.765}{$u_{1000}$}

            \labelbox{text=red}{0.1, 0.585}{$v_1$}
            \labelbox{text=red}{0.34, 0.585}{$v_2$}
            \labelbox{text=red}{0.585, 0.585}{$v_3$}
            \labelbox{text=red}{0.825, 0.585}{$v_{10}$}
            \labelbox{text=red}{0.10, 0.47}{$v_{50}$}
            \labelbox{text=red}{0.34, 0.47}{$v_{250}$}
            \labelbox{text=red}{0.59, 0.47}{$v_{500}$}
            \labelbox{text=red}{0.835, 0.47}{$v_{1000}$}

            \node[rotate=-16, font=\normalsize] at (0.88, 0.24) {$k^{-5/3}$};

        \end{scope}
    \end{tikzpicture}
    \caption{Top two rows: Eight representative spatial SVD modes ($u_i$) and temporal modes ($v_i$) of the decomposed horizontal velocity component $u$ at the surface for the S2 
\add{
    data
}%
    case, chosen for illustrative purposes, for a snapshot at an arbitrary point in time. Middle two rows: Evolution of the same modes in time (represented by frame number). Lower left: singular values of the SVD modes%
\akr{
    , with inset showing the first 350 values%
}%
    . Lower right: Normalized turbulent power-density spectrum (PSD) for velocity fields compressed by retaining only modes from $1$ to rank $r$.}
    \label{fig:SVD_DNS}
\end{figure}

\begin{figure}[!htbp]
    \centering
    \begin{tikzpicture}
        \node[anchor=south west, inner sep=0] (image) at (0,0) {
            \includegraphics[width=\textwidth, trim={0.1cm 0cm 0cm 0cm}, clip]{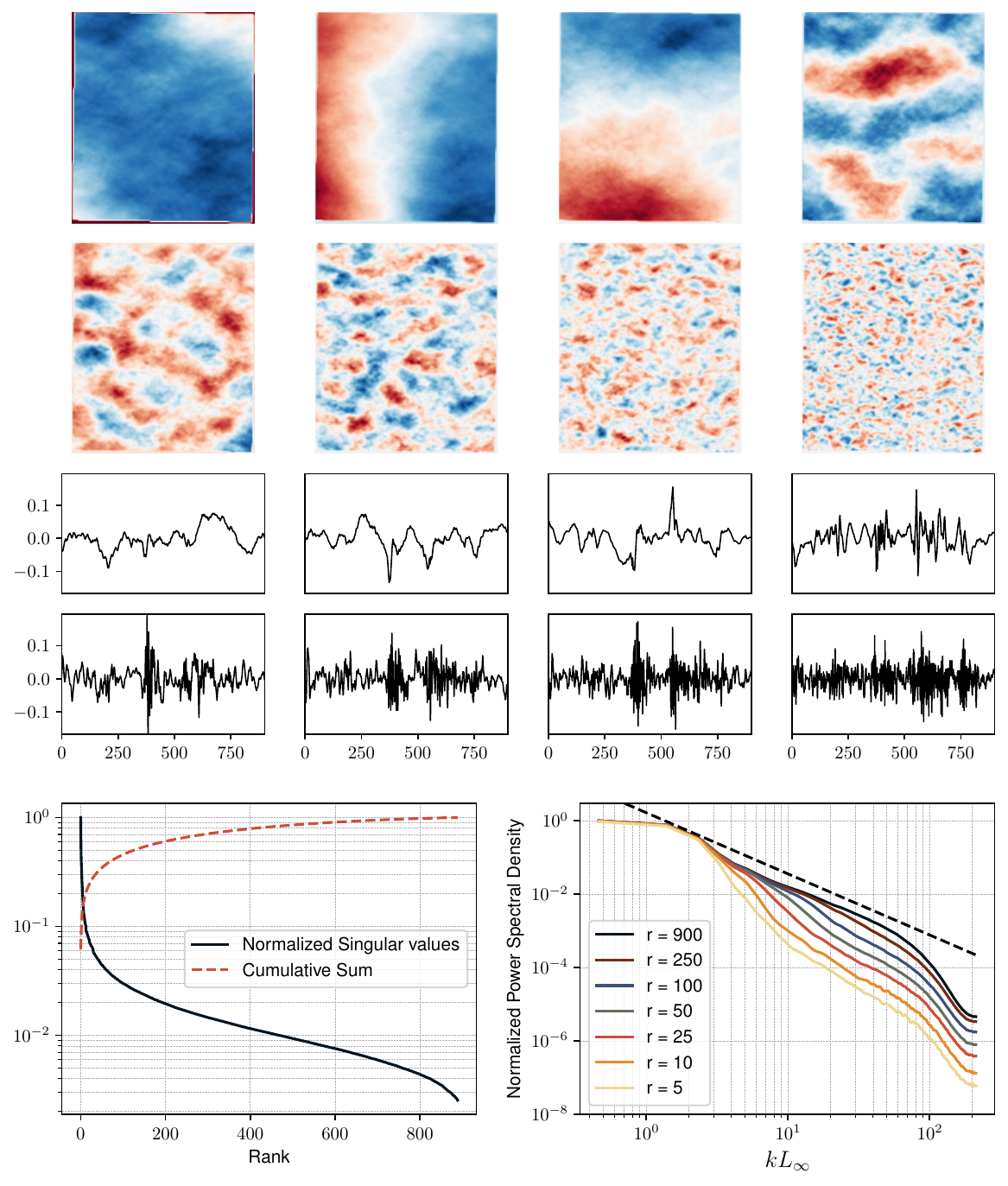}
        };

        \begin{scope}[shift={(image.south west)}, x={(image.south east)}, y={(image.north west)}]
 
            \labelbox{}{0.10, 0.96}{$u_1$}
            \labelbox{}{0.34, 0.96}{$u_2$}
            \labelbox{}{0.585, 0.96}{$u_3$}
            \labelbox{}{0.825, 0.96}{$u_{10}$}
            \labelbox{}{0.10, 0.765}{$u_{25}$}
            \labelbox{}{0.34, 0.765}{$u_{50}$}
            \labelbox{}{0.59, 0.765}{$u_{100}$}
            \labelbox{}{0.835, 0.765}{$u_{250}$}

            \labelbox{text=red}{0.1, 0.585}{$v_1$}
            \labelbox{text=red}{0.34, 0.585}{$v_2$}
            \labelbox{text=red}{0.585, 0.585}{$v_3$}
            \labelbox{text=red}{0.825, 0.585}{$v_{10}$}
            \labelbox{text=red}{0.10, 0.47}{$v_{25}$}
            \labelbox{text=red}{0.34, 0.47}{$v_{50}$}
            \labelbox{text=red}{0.59, 0.47}{$v_{100}$}
            \labelbox{text=red}{0.835, 0.47}{$v_{250}$}

            \node[rotate=-16, font=\normalsize] at (0.88, 0.245) {$k^{-5/3}$};

        \end{scope}
    \end{tikzpicture}
    \caption{Same as Fig.\ \ref{fig:SVD_DNS}, but for the 
    \rjo{turbulence
    }{high--Reynolds number}
    experimental 
    \ajo{
    dataset
    }%
    E2. 
    \djo{case.}
    SVD modes are found from 
    \rjo{
    the turbulent streamwise velocity component
    }{a horizontal velocity component measured in a plane located}
    1\,cm ($0.14 L_{\infty}$) below the free surface. Note that the total number of SVD modes is 900%
    \replace{in the experimental case.}{
    here.
    }
    }
    \label{fig:SVD_Exp}
\end{figure}

\rjo{
Figs.\
}{Figures} 
\ref{fig:SVD_DNS} and \ref{fig:SVD_Exp} show the SVD modes and turbulence power spectra for 
\replace{cases }{
datasets
}%
S2 (DNS) and E2 (experiment), respectively. Through SVD the flow data are decomposed in matrices that contain spatial ($\bU$), temporal ($\bV$), and energy ($\bS$) information (strictly speaking, the latter matrix contains the singular values, $\sigma_i$, which can be considered as a measure of energy for the flow patterns in $\bU$ and $\bV$). Decomposing a dataset of $n$ timesteps results in $n$ modes. For example, 
\replace{case}{
dataset
}%
S2 has $12500$ modes; E2 has $900$. The modes are ordered by energy/singular-value, hence $\sigma_1 \geq \sigma_2 \geq \dots \geq \sigma_n$.  After decomposition, the exact flow dataset can be reconstructed by matrix multiplication of $\bU, \bS$ and \ajo{
$\bV^T$%
}%
.
Spatial and temporal coefficients of eight modes for a velocity field near the surface are shown in \rjo{the upper and middle panels of the }{rows 1--2 (spatial) and 3--4 (temporal), in }\rjo{f}{F}igs.\ \ref{fig:SVD_DNS} and \ref{fig:SVD_Exp}. We observe that the large-scale spatial modes have large singular values, meaning the majority of the energy content is found in these modes---as expected from our knowledge of the turbulent flow. Likewise, modes of higher order (smaller singular value/lower energy) are related to smaller structures, as seen from comparing, e.g., $u_2, \sigma_2$ to $u_{50}, \sigma_{50}$ in Fig.\ \ref{fig:SVD_DNS}.  Moreover, we observe a more rapid variation in the temporal coefficients as the order is increased. This is as expected for small compared to large scales in the turbulence, yet it may also be a result of noise, which typically shows up in small-scale data, another argument for using a compressed rather than a full dataset with SHRED.

The choice of rank (i.e., the range of modes to retain) for the 
\ajo{%
compressed,
}%
low-rank representation of the data is done on the basis of 1) the observance of the spatial and temporal modes and their scales and variance, 2) the rapid decline of singular values per rank number, 3) from the turbulence \djo{energy }spectra dependency on the rank number, and 4) post-result evaluation of optimal values for best SHRED performance. 
From a computational perspective (points 1 \& 2), the optimal low-rank representation contains the relevant modes and covers most of the cumulative sum of the singular values. From a fluid mechanics perspective (point 3), we want to choose a rank \ajo{
of
}%
truncation that 
\ajo{
yields a compressed dataset which
}%
matches the original spectrum well into the upper part of the inertial range.
The results in Figs. \ref{fig:SVD_DNS} and \ref{fig:SVD_Exp} suggest that rank 250 is sufficient for the S2 
\replace{case}{
dataset%
}%
, and rank 100 is sufficient for E2. This corresponds to a cutoff at normalized wavenumber $k_cL_{\infty}$ 
\akr{
roughly equal to $3$--$10$, where $k_c$ is the wavenumber where the 1D turbulence spectra of the compressed and uncompressed data first deviate by more than 10\% (see \S \ref{subsec:PSD})%
}%
. Hence, all scales down to a 
\akr{%
third
}%
of the integral length scale \akr{%
for DNS,
and sixth to tenth for experiments,
}%
\dkr{,} are resolved. A parametric study of SHRED performance for different 
\rjo{
rank truncations
}{levels of compression} 
is included in the appendix \ref{sec:app1}. We find that the performance of SHRED decreases 
\rjo{
past
}{when the rank we set for compression is higher than}
a certain 
\djo{
rank
}%
value 
\ajo{
i.e., when the level of compression is too low
}%
, with the exact value 
\replace{depending on the flow case. }{
being different for each flow.
}%
Taking this into account, we choose the low-rank representation of the data to be truncated as listed in Table \ref{tab:SVD_rank}. We note that for experimental 
\add{
data
}%
cases, we choose a rather low rank, with only a cumulative sum of singular values of $45$ to $50$ percent. This is because the noise present in the experimental data makes it significantly harder to handle higher-rank data
    \delete{in these cases }%
than in the data from the DNS.   

\begin{table}
    \begin{center}
    \begin{tabular}{c c c c c c}

      \hspace{0.4em} Dataset \hspace{0.4em} & \hspace{0.4em} Full-rank \hspace{0.4em} & \hspace{0.4em} Low-rank \hspace{0.4em} & \hspace{0.4em} Cumulative sum \hspace{0.4em} & \hspace{0.4em} Rank truncation \hspace{0.4em} & $k_cL_{\infty}$ \\
        \hline
    S1  &   10900& 225  & 75.0\%  & 97.9 \%   & \akr{3.6} \\ 
    S2  &  12500 & 250  & 72.0\%  & 98.0 \%   & \akr{3.7}  \\ 
    E1  &  900 & 100  & 52.0\%  & 89.9 \%  & \akr{6.9} \\ 
    E2  &  900 & 100  & 45.0\%  & 89.9\%   & \akr{10.3} \\
    \end{tabular}
         \caption{
\rjo{Chosen low-rank truncation numbers (i.e. number of SVD modes included in the compressed dataset) for the SVD compression of the different datasets including the cumulative sum of singular values in the compressed data, and SVD mode truncation, i.e. low-to-full rank number ratio. We also include
}{SVD compression details. From left: Dataset, number of SVD-modes in the uncompressed data, number of SVD-modes in the compressed data, normalized cumulative sum of singular values of the compressed data, percentage of modes left out of the compressed dataset, and }
         the upper cutoff wavenumber 
\djo{for }%
         where low- and full-rank turbulence spectra start to deviate by more than 10\%, normalized by integral length scale.}
    \label{tab:SVD_rank}
    \end{center}
\end{table}

\subsection{Performance metrics}
\label{sec:method_metrics}

When evaluating SHRED performance, we choose a set of 
\ajo{
depth-dependent error
}%
metrics that capture different aspects of the reconstruction error. 
\ajo{
When reporting results in terms of these metrics in \S \ref{sec:results}, we set ``ground truth" to be the uncompressed, full-rank
}\djo{
We choose to measure every metric relative to the full-rank ground truth
}%
velocity fields, rather than the compressed data, 
\ajo{
although SHRED is trained on the latter. Behind this choice is the fact that our aim in application of SHRED to real-world flows is to accurately capture the full flow fields, not their compressed version. To explore the limitations on SHRED due to the compression step, we nevertheless include versions of these error metrics where the reconstructions are measured relative to the compressed data as ``ground truth" in Appendix \ref{sec:app1}.
}\djo{as the former are the fields of interest for reconstruction in practice.} 

\subsubsection{Time-averaged velocity profile}
The time-averaged velocity profile of a free-surface turbulent flow gives a sense of the effect of the upper boundary on the subsurface flow. It is essentially a measure of the root-mean-square velocity component(s) averaged over all snapshots, computed for component $u$ by
\begin{equation}
    \langle\urms\rangle = \left\langle\left(\frac1{mn}\sum_{x=1}^m \sum_{y=1}^n u_{x,y}^2 \right)^\frac12\right\rangle \, ,
    \label{eq:time_avg_urms}
\end{equation}
where $\langle\cdots\rangle$ denotes averaging in time (i.e., over all snapshots) and spatial averaging is performed as detailed in connection with Eq.\ \eqref{eq:urms}. Unlike the metrics presented below, the time-averaged velocity profile is not a performance metric \emph{per se}, but computing and comparing such profiles for ground truth data and reconstructed data enable a straightforward assessment of reconstruction performance.

\subsubsection{Normalized mean squared error}
Normalized mean squared error (NMSE) is one of the simplest and most widely used metrics for reconstruction accuracy. We use the normalized form:
\begin{equation}
    \NMSE = \frac{\left < \sum_{x=1}^m \sum_{y=1}^n |\tilde u_{x,y} - u_{x,y}|^2 \right >}{\left < \sum_{x=1}^m \sum_{y=1}^n u_{x,y}^2 \right >} \, ,
    \label{eq:NMSE}
\end{equation}
where $\tilde u$ is the reconstruction field and $u$ is the ground truth.
Hence, the MSE is calculated as the spatial (over each plane) and temporal mean of the square of the difference between the ground truth and reconstruction, and normalized by the mean square of the ground truth. The MSE is a simple measure of the error in the values themselves, at either a single point or, as in this case, the averaged error of data points over a plane. The MSE captures the loss in amplitudes in the reconstruction. However, it has drawbacks such as not taking into account the sign of the signal, disregarding spatial structures, and treating each data point as equally important \cite{Wang2009_signal_metrics} as we discuss further below.

\subsubsection{Power spectral density error}\label{subsec:PSD}

A common description of turbulence is to regard the power spectral density (PSD). Given snapshots $u(x,y,t_n)$, we define the 1-D PSD 
\ajo{
for the single velocity component $u$
}%
along $x$, averaged over $y$ and time, as
\begin{equation}
  E(k)
  \;=\;
  \frac{1}{N_t\,N_y}\sum_{n=1}^{N_t}\sum_{j=1}^{N_y}
  \Big|\;\mathcal{F}_x\!\big[w(x)\,u(x,y_j,t_n)\big](k)\;\Big|^2,
  \label{eq:psd_def}
\end{equation}
where $\mathcal{F}_x$ denotes the discrete Fourier transform in $x$, $N_t$ and $N_y$ are the number of time steps and grid points in y-direction respectively, and the Hanning window function is denoted by $w(x)$.

To compare the different flow cases, 
\rjo{a normalized PSD (NPSD) is used, which we}{
we make use of a normalized 1D spectrum,
}%
defined as 
\begin{equation}
  \mathrm{NPSD}(k)
  \;=\;
  \frac{E(k)}{\displaystyle \max_{k}\,E(k)}\,,
\end{equation}
i.e. spectra are normalized by the maximum of the ground truth \ajo{1D} spectrum 
\replace{for that }{
in each
}%
case. 
\djo{The PSD provides a measure of how energy is distributed across spatial length scales.}

In order to evaluate the \rjo{energy spectrum of the reconstructed turbulent velocity field as compared to ground truth}{reconstruction accuracy in terms of the 1D PSD of the velocity field $u$}, we introduce the 1D power spectral density error (PSDE) as the relative error in integrated 
\rjo{
spectral energy
}{spectrum}
up to a cutoff wavenumber $k_c$
\ajo{
(details below and Table \ref{tab:SVD_rank})%
}%
, defined as
\begin{equation}
  \mathrm{PSDE}
  \;=\;
  \frac{\Big|\displaystyle\int_{0}^{k_c}\! \tilde{E}(k)\,\mathrm{d}k
              \;-\;\displaystyle\int_{0}^{k_c}\!E(k)\,\mathrm{d}k\Big|}
       {\displaystyle\int_{0}^{k_c}\!E(k)\,\mathrm{d}k}\, ,
  \label{eq:PSDE}
\end{equation} 
where $\tilde{E}(k)$ and $E(k)$ are the 
\rjo{
PSDs
}{1D spectra}
for the reconstruction and ground truth 
\rjo{
spectra
}{data,}
respectively, 
\rjo{
given
}{computed} 
by Eq.~\eqref{eq:psd_def}. In the discrete implementation we use Simpson’s rule on the FFT wavenumber bins $k_i$ with weights $w_i$, and the PSDE thus becomes
\begin{equation}
  \mathrm{PSDE}
  \;=\;
  \frac{\left|\sum_{i=0}^{i_c} w_i\,\tilde{E}_i \;-\; \sum_{i=0}^{i_c} w_i\,E_i\right|}
       {\sum_{i=0}^{i_c} w_i\,E_i}\,,
\end{equation}
where $i_c$ is the index for the wavenumber bin $k_c$. 

We choose 
\ajo{the cutoff wavenumber
}%
$k_c$ to exclude scales that are not reliably represented by the low-rank data used in
training. Operationally, $k_c$ is set near the onset where the rank-truncated SVD spectrum begins to
depart noticeably ($\approx10\%$) from the ground-truth spectrum; in practice this falls in the
intermediate (inertial-range) wavenumbers for all datasets. The same $k_c$ is used for ground truth
and reconstructions within a given 
\replace{case. }{
dataset.
}%
By setting these integration limits, rather than using the full spectrum, we neglect the major contribution in the error (when compared to the full-rank ground truth) the low-rank SVD truncation process itself would produce. Therefore, it is only relevant to compare the spectral power of the SHRED reconstruction and full rank ground truth down to the low-rank SVD-resolved spatial scales. For the experimental data, this corresponds to a wavelength resolution of $\lambda_c = 2\pi /k_c = 1.5$ cm, which is about 
\rkr{25\%}{
20 \% (E1) and 13 \% (E2)
}%
of the integral length scales. For the DNS data, the cutoff wavelength is chosen at around 
\rkr{17\%}{
11\%
}%
of the integral length scale.

\subsubsection{Structural similarity index measure}

The Structural Similarity Index Measure (SSIM) is a metric often used to compare image quality in terms of visual perception \cite{SSIM_Zhou2004}. While MSE captures the point-by-point local mismatch of the average amplitude of an image, SSIM also emphasizes structural correlation and contrast, outperforming MSE and PSNR (see below) in evaluating for visual similarity as described, e.g., by Wang \& Bovik \cite{Wang2009_signal_metrics}. It is designed to be more aligned with human visual perception, and is therefore often used in computer vision and deep learning.
The SSIM we use is defined as
\begin{equation}
    \SSIM = l(\tilde u,u) c(\tilde u,u) s(\tilde u,u),
    \label{eq:ssim}
\end{equation}
where $\tilde u$ is our reconstruction image, $u$ is the full-rank ground truth, $l(\tilde u,u)$ is the luminance similarity factor, $c(\tilde u,u)$ the contrast similarity factor, and $s(\tilde u,u)$ is the structure similarity factor, all defined and discussed in \cite{SSIM_Zhou2004}. 
The SSIM can take values between $-1$ (anti-correlated) and $1$ (structurally identical), with $0$ indicating no similarity. 

\subsubsection{Peak signal-to-noise ratio}
The peak signal-to-noise ratio (PSNR) is a measure of image resolution widely used in computer vision and reconstruction tasks.
Essentially, it quantifies how much of the signal in an image is relevant compared to irrelevant noise. It is inversely related to the MSE metric, and is defined as
\begin{equation}
    \PSNR = 10 \log_{10}\left( \frac{\MAX_j^2}{\NMSE} \right),
    \label{eq:PSNR}
\end{equation}
where $\MAX_j$ is the maximum pixel bit value of image $j$, and $\NMSE$ is given as in Eq.\ (\ref{eq:NMSE}). Generally, the more noisy and distorted the image, the lower the PSNR value. The metric is usually given in decibels (dB), and typical values for moderate quality images are 20-30, while exceptionally good quality images will show values of 30-50 \cite{SSIM_Zhou2004}. Although it is closely related to the MSE, we include PSNR as a metric because it accounts for the maximal value in an image, creating a reference point for the signal-to-noise ratio. 

\section{Results and discussion}
\label{sec:results}
In what follows, we present and discuss the results of reconstructing subsurface turbulent velocity fields from sparse measurements of the free surface only, by applying the SHRED algorithm to 
\replace{our }{
the
}
turbulence data from the DNS 
\add{
\cite{aarnes2025b}
}
and experimental 
\add{
\cite{babiker2026}
}
\replace{cases. }{
datasets.
}%
We first show and discuss the field reconstructions for shallow and deep horizontal planes, and compare these to the 
\rjo{
original
}{uncompressed} 
and compressed fields. Secondly, we present a detailed analysis of the depth-dependent performance of SHRED, using the set of error metrics presented in 
Sec.\ \ref{sec:method_metrics}. Third, 
\ajo{
we compare the performance of SHRED with a proper orthogonal decomposition based method for sensor-to-flow mapping. Forth, 
}%
we demonstrate the capabilities of SHRED in temporal dynamics by looking at the RMS velocity time series of reconstructed fields. Finally, we show that the reconstructed fields yield reasonable turbulent power spectra for the relevant spatial scales, as compared to the compressed and ground-truth fields.  

\subsection{Reconstruction of surface elevation and velocity fields}

\begin{figure}[!htbp]
\centering

\begin{minipage}{0.91\textwidth}
\centering
\begin{overpic}[width=\textwidth, trim={1cm 1cm 1cm 0cm}, clip]
    {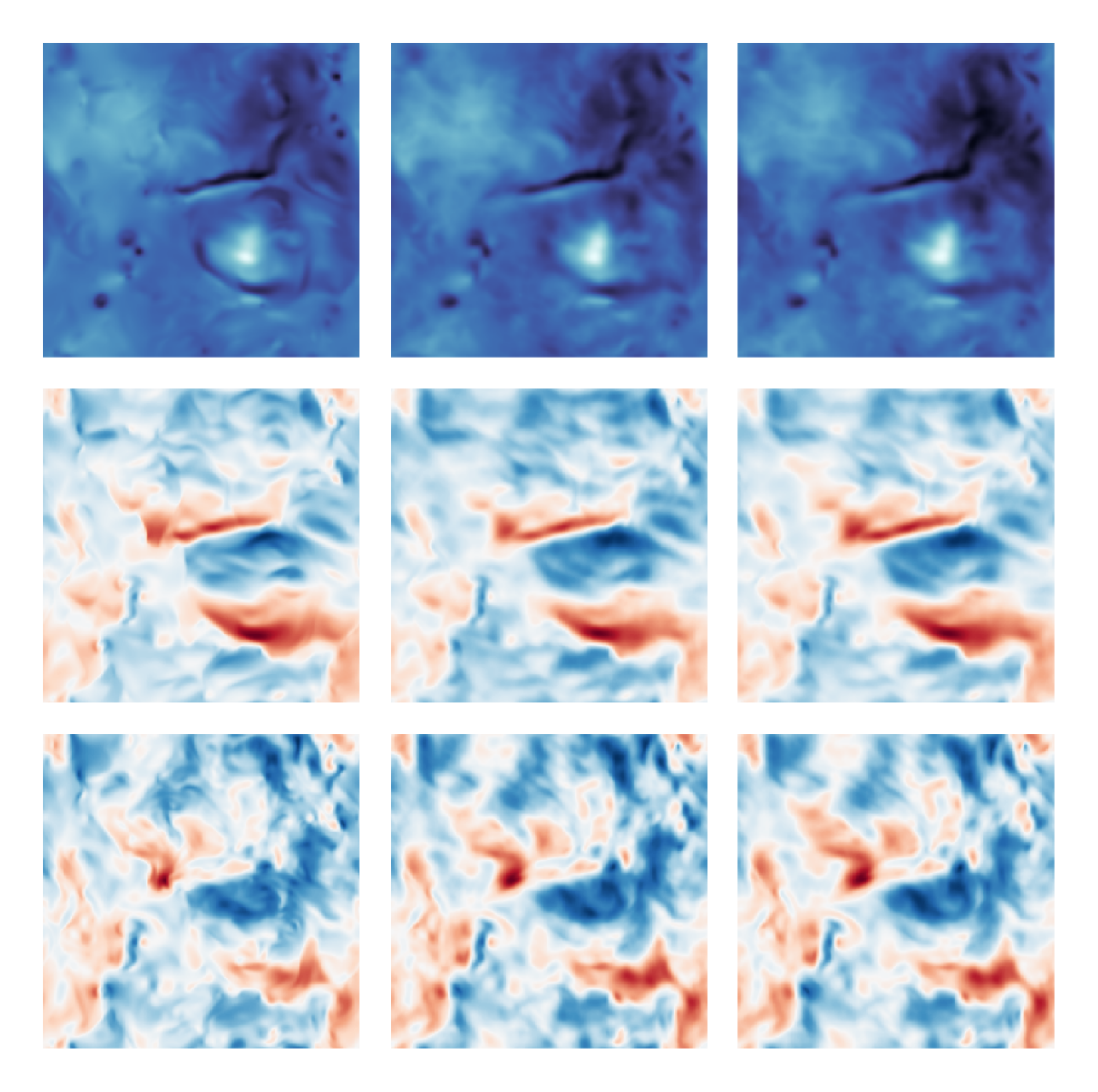}
    \put(0,0){%
    \begin{tikzpicture}[x=1bp,y=1bp]        

    \useasboundingbox (0,0) rectangle (400,500);

    \labelbox{fill opacity=0.6, anchor=west}{10, 405}{Surface elevation}
    \labelbox{fill opacity=0.6, anchor=west}{10, 265}{Surface velocity}
    \labelbox{fill opacity=0.6, anchor=west}{10, 125}{Bulk velocity}

    \node[anchor=center, font=\normalsize] at (65,430) {Ground truth};
    \node[anchor=center, font=\normalsize] at (210,430) {Compressed};
    \node[anchor=center, font=\normalsize] at (352,430) {Reconstruction};

    \end{tikzpicture}%
    }
\end{overpic}
\end{minipage}
\hfill
\begin{minipage}{0.08\textwidth}
\centering

\vspace{1.5cm}
\includegraphics[width=\linewidth]{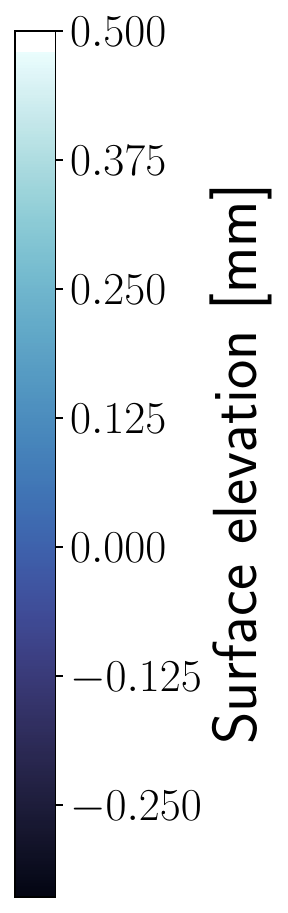}
\vspace{0.0cm}

\includegraphics[width=\linewidth]{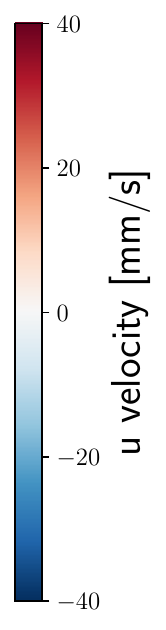}
\vspace{0.4cm}
\includegraphics[width=\linewidth]{figures/u_color_bar_S2.pdf}

\end{minipage}

\caption{Demonstration of the reconstruction capabilities of SHRED for 
    \replace{the DNS case S2: }{
    DNS dataset S2:
    }%
    Ground truth (left-hand column), compressed training data with rank $250$ (middle column) and SHRED reconstructions of the compressed data (right-hand column), for surface elevation (top) and horizontal velocity component $u$ near the surface at depth $0.02 L_{\infty}$ (middle row) and in the bulk at depth $1.2 L_{\infty}$(bottom). In the top row, light (dark) colors illustrate elevation (depression) of the surface compared to the mean level; in the middle and bottom rows darker red (blue) indicates higher positive (negative) values of the velocity $u$ with white representing zero velocity. The size of each plotted region is approximately $6L_\infty \times 6L_\infty$.}
\label{fig:reconstruction_RE2500}

\end{figure}

\begin{figure}[!htbp]
    \begin{subfigure}[b]{\textwidth}
         \centering
          \begin{overpic}[width=\textwidth, trim={1cm 1cm 1cm 1cm}, clip, grid=off, unit=2bp,tics=2]{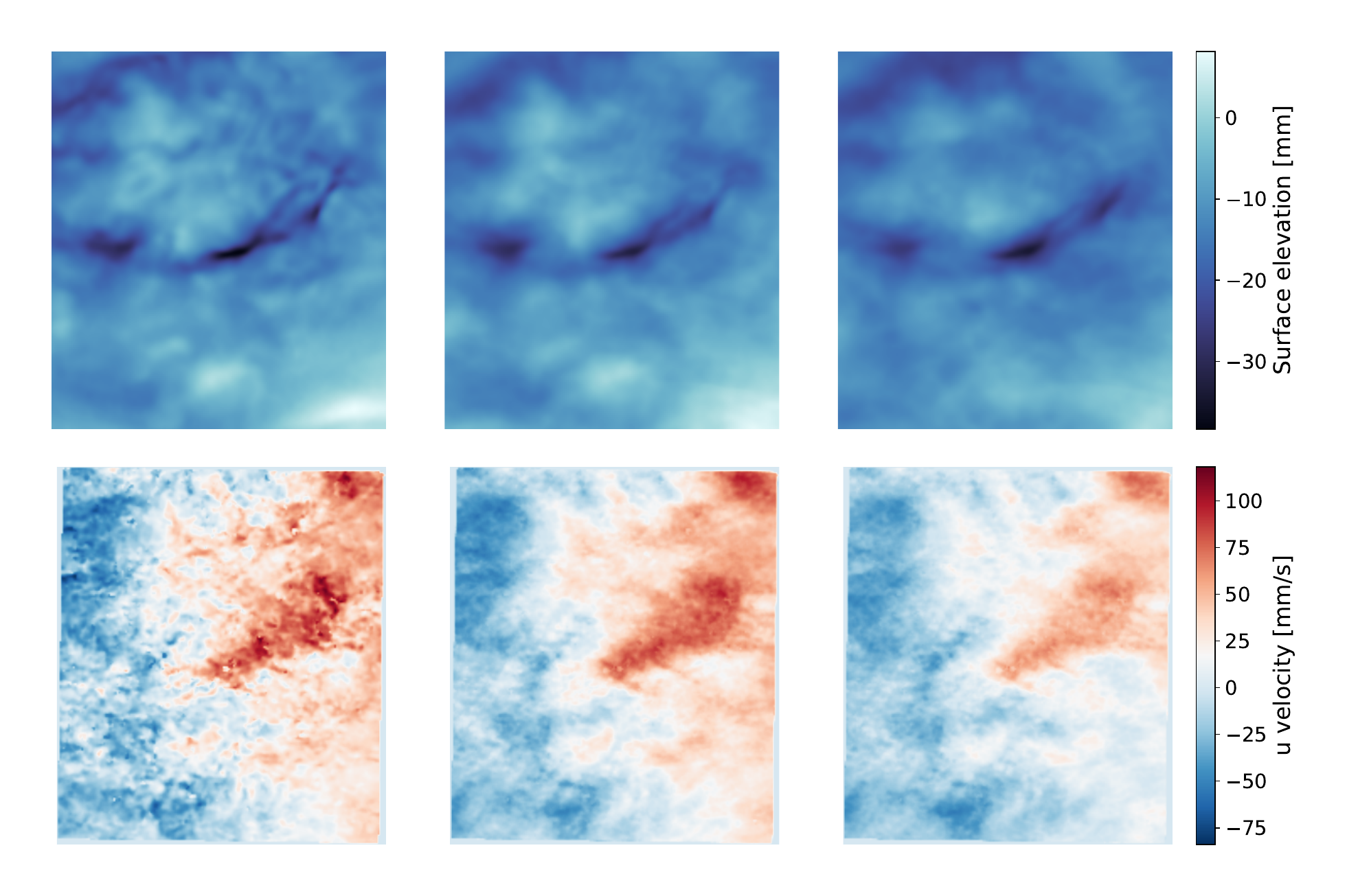}
            \put(0,0){%
            \begin{tikzpicture}[x=1bp,y=1bp]        
       
            \useasboundingbox (0,0) rectangle (400,350);

            \labelbox{fill opacity=0.6, anchor=west}{10, 289}{Surface elevation}
            \labelbox{fill opacity=0.6, anchor=west}{10, 135}{subsurface velocity,}
            \labelbox{fill opacity=0.6, anchor=west}{10, 120}{Depth 1 cm}

            \node[anchor=center, font=\normalsize] at (60,307) {Ground truth};
            \node[anchor=center, font=\normalsize] at (210,307) {Compressed};
            \node[anchor=center, font=\normalsize] at (355,307) {Reconstruction};
            \end{tikzpicture}%
            }
            \end{overpic}
         \label{fig:reconstruction 1cm}
     \end{subfigure}
     
     \begin{subfigure}[b]{\textwidth}
         \centering
            \begin{overpic}[width=\textwidth, trim={1cm 1cm 1cm 11.5cm}, clip, grid=off, unit=2bp,tics=2]{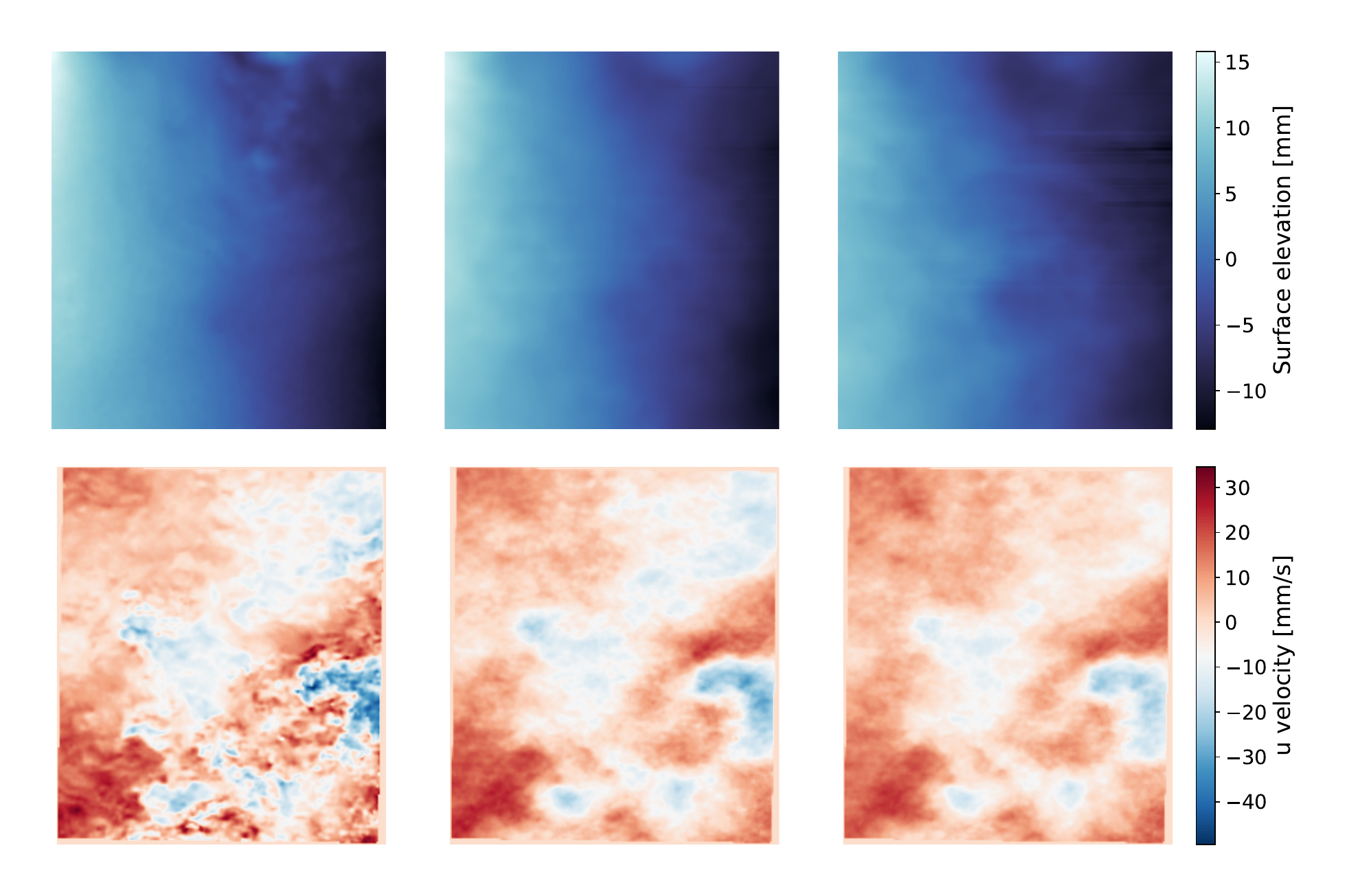}
            \put(0,0){%
            \begin{tikzpicture}[x=1bp,y=1bp]        
       
            \useasboundingbox (0,0) rectangle (400,350);

            \labelbox{fill opacity=0.6, anchor=west}{10, 135}{subsurface velocity,}
            \labelbox{fill opacity=0.6, anchor=west}{10, 120}{Depth 10 cm}
            \end{tikzpicture}%
            }
            \end{overpic}

         \label{fig:reconstruction 10 cm}
     \end{subfigure}
    \caption{
    Same as Fig.\ \ref{fig:reconstruction_RE2500} but for experimental
    \add{
    data, flow
    }%
    case E2: Ground truth (left-hand column) compared to compressed training data of rank $100$ (middle column) and SHRED reconstructions of the compressed data (right-hand column), for surface elevation (top) and the corresponding horizontal velocity component $u$, 1 cm \akr{($0.10 L_{\infty}$)} and 10 cm \akr{($0.85 L_{\infty}$)} beneath the surface, in the middle and bottom rows, respectively. The surface elevation is simultaneous with the velocity in the middle row, whereas measurements in the deeper plane were taken at a different time; their corresponding surface elevation is not shown here.
    The size of the surface elevation field is \akr{$1.86 L_{\infty} \times 2.1 L_{\infty}$}, and slightly smaller at \akr{$1.68 L_{\infty} \times 1.92 L_{\infty}$} for the velocity fields.} 
    \label{fig:reconstruction Exp}
\end{figure}

To illustrate the performance of SHRED qualitatively, we present a side-by-side comparison of the surface elevation and a component of the horizontal velocity field: the original, uncompressed data, their low-rank SVD approximations, and the SHRED reconstructions, for the S2 DNS 
\replace{case }{
data
}
in Fig.\ \ref{fig:reconstruction_RE2500} and the experimental 
\rsm{data }{
case
}%
E2 in Fig.\ \ref{fig:reconstruction Exp}.
The immediate eyeball observation is that the fields lose some sharpness in the SVD compression step, whereas the compressed and reconstructed fields are only distinguishable by eye upon careful inspection.

From Fig.\ \ref{fig:reconstruction_RE2500} one might get the impression that errors in the final reconstruction of the fields (right-hand column) occur during compression before training while the subsequent steps reproduce the compressed fields near-perfectly. To some extent, this is correct. However, increasing the number of SVD modes in compression beyond the ranks chosen does not necessarily improve the reconstruction. As we detail in Appendix \ref{sec:app1}, training SHRED using the original uncompressed dataset or with very little compression leads to considerable overfitting and far poorer results, as well as higher computational cost.

Inspecting the surface elevation for DNS case S2 in the top row of Fig.\ \ref{fig:reconstruction_RE2500}, we notice a sharp, dark, depression curve --- referred to as a ``scar'' \cite{brocchini2001} --- and brighter areas which are ``boils'' which signify upwelling of fluid to the surface. These features are successfully reconstructed, although the amplitudes can be seen to have been somewhat dampened. Scars also manifest in the surface velocity field
\delete{of case S2, }
(middle row) as areas of fast flow. As with the surface elevation, the reconstruction looks visually indistinguishable from the low-rank approximation and the ground truth. Moreover, as demonstrated in the bottom row in Fig. \ref{fig:reconstruction_RE2500}, SHRED is capable of reconstructing the velocity field in the bulk flow, that is, at depths below the surface-influenced layer (depth $\bar{z} \gtrsim L_\infty$ for case S2; details on the surface-influenced layer in \cite{aarnes2025}).

Similarly, for the experimental 
\replace{case }{
dataset
}%
E2, Fig.\ \ref{fig:reconstruction Exp} shows visually good reconstruction performances across several planes. It is a considerably more challenging task to reconstruct planes in the experimental 
\replace{cases, }{
flows
}%
due to noise and the much greater disparity of time and length scales at the far higher turbulent Reynolds numbers. However, we observe that even in this case, SHRED is capable of reconstructing from the time series of surface elevation in only three points, large and strong features such as scars at the surface elevation and medium-sized structures in the velocity fields as far down as $10$\,cm ($z\approx L_\infty$) below the surface, although smaller features are somewhat blurred. More clearly than in the DNS 
\replace{case, }{
flow in Fig.\ \ref{fig:reconstruction_RE2500},
}
velocity amplitudes are generally diminished in the reconstruction (right column) compared to the compressed training data (middle column).

\rjo{As an eyeball metric, Figs.}{
Figures
}%
\ref{fig:reconstruction_RE2500} and \ref{fig:reconstruction Exp} illustrate that SHRED is capable of 
\rjo{
reconstructing
}{qualitative reconstruction of}
flow fields from sparse height measurements at the free surface in free-surface 
\replace{flows in both both DNS simulations and experimental cases.}{
turbulence data from DNS as well as state-of-the-art experiments.
}%
SHRED's performance is somewhat weaker in the experimental case, which is to be expected due to the presence of noise, greater range of turbulent scales, and significantly less data available for training compared to the long, continuous DNS datasets. In light of this we find it remarkable that SHRED reconstructs features of centimeter size $10$\,cm beneath the surface, well 
\replace{into the deeper part of}{ 
outside the
}%
``blockage layer'' where the free surface directly influences the velocity field \cite{shen99,aarnes2025}.

\FloatBarrier

\subsection{Depth-dependent performance}
\label{sec:depth_error_results}

To quantify the performance of SHRED beyond simple comparisons, we use the five metrics discussed in Sec.\ \ref{sec:method_metrics} to evaluate different aspects of SHRED's reconstruction of the horizontal velocity field $u$. 
\akr{As stated in \S \ref{sec:method_metrics}, we report all error metrics relative to the original, uncompressed fields unless otherwise stated. The primary objective is to reconstruct the true flow dynamics, with compression serving solely as an intermediate representation to facilitate optimal learning. A detailed analysis of the effect of the SVD compression on the error is provided in Appendix~\ref{sec:app1}.
}%
In addition to the averages taken in horizontal space and time, we perform ensemble averaging over $25$ individual reconstructions for each flow case, where every 
\replace{ensemble case }{
time series in the ensemble
}%
is constructed by a random distribution of the full dataset into training, validation, and testing. 
The results are displayed with depth $z$ scaled by the integral length scale on the ordinate axis. Note that we adopt the convention from Aarnes et al.~\cite{aarnes2025} of using an average measure for `horizontal' grid plane depth for the DNS data, to allow for straightforward comparison of flow variables at a grid plane on the undulating grid without interpolation (details in \cite{aarnes2025}).

\begin{figure}[!htbp]
    \centering
    
    \begin{overpic}[width=0.98\textwidth, trim={0cm 0cm 0cm 0.1cm}, clip, grid=off, unit=2bp,tics=2]{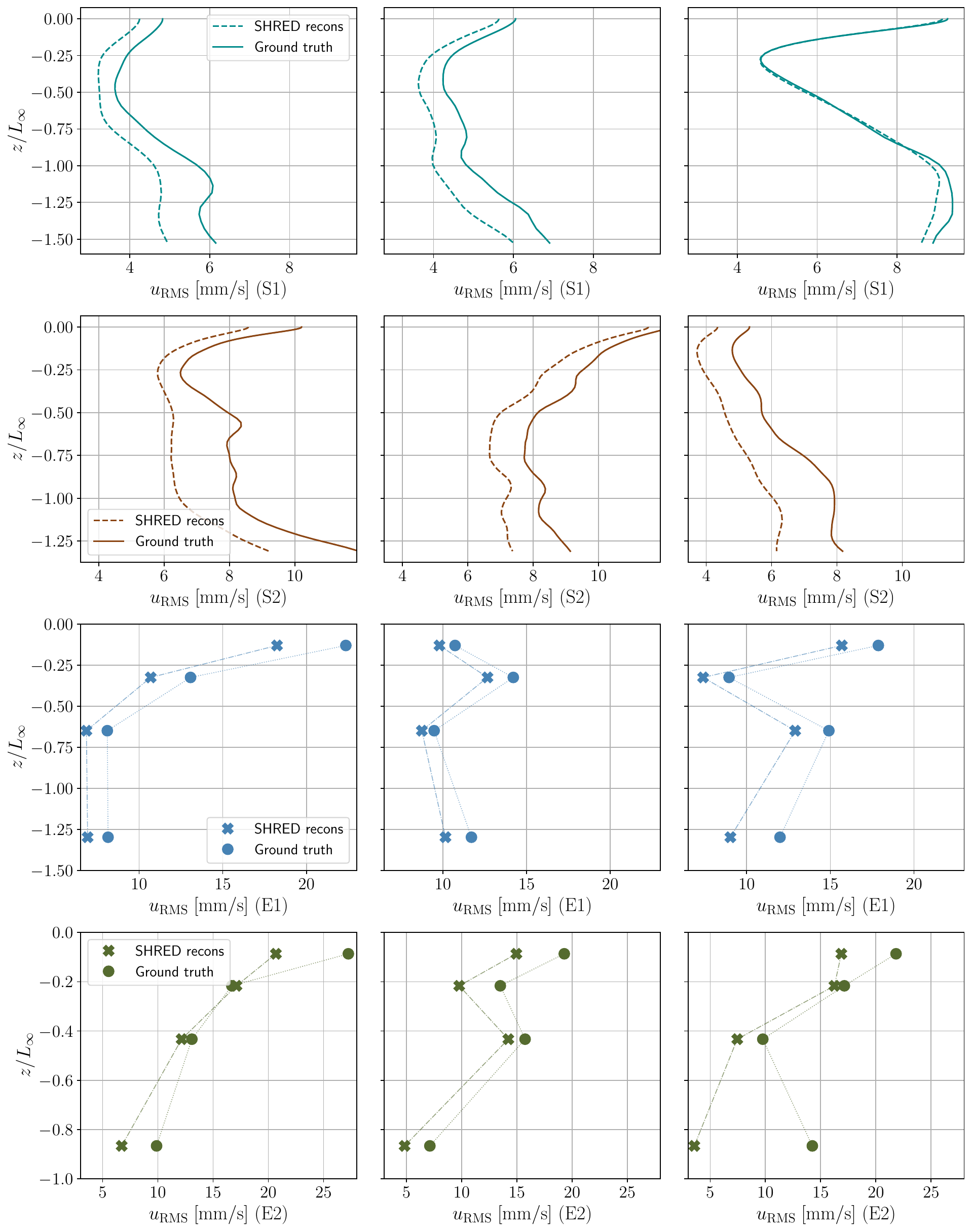}
    \put (18,100) {\makebox[0pt]{\Centerstack{Test instant 1}}}
    \put (43,100) {\makebox[0pt]{\Centerstack{Test instant 2}}}
    \put (67,100) {\makebox[0pt]{\Centerstack{Test instant 3}}}
    \end{overpic}
    \caption{Examples of instantaneous profiles of $\urms(z,t)$ for ground truth (solid lines) and reconstruction (dashed lines) for the four \replace{cases }{
    datasets
    }%
    S1, S2, E1 and E2 (rows from top to bottom) at three randomly chosen time instants (columns from left to right). 
    \akr{
    Note that for S1 \& S2, the vertical profile per dataset and test instant is simultaneous, while for E1 \& E2 the $\urms$ at each plane is measured independently from other planes, i.e. vertical profile is not simultaneous, but rather a combination of 4 separate instants, each with their separate SHRED model.
    }%
    }
    \label{fig:rms_instantaneous}
\end{figure}

Because the error metrics are averaged over time and ensembles, which could potentially conceal the direct performance error image-by-image, we 
\rjo{
include
}{first consider}
instantaneous profiles of $\urms(z,t)$ in Fig. \ref{fig:rms_instantaneous}, comparing reconstructions (dashed lines) to the uncompressed ground-truth values (solid lines) at three different arbitrarily chosen time instants. 
Note that for the two experimental 
\replace{cases, }{
flows,
}%
the 
\ajo{
subsurface velocity
}%
measurements taken at different 
\rjo{
values of $z$
}{depths $z$}
are not taken at the same time since different planes were measured separately.
Generally, the reconstruction profiles match the ground truth profiles in shape but for a drop in magnitude, different for different depths and instants. There is a tendency for rapid changes in the measured profiles to be smoothed out in the reconstructed ones, particularly visible here for 
\replace{case }{
flow
}%
S1 at instant 1. Such rapid changes are typically caused by intermittent turbulent events which are difficult for a neural network to learn, in particular for time-series-based networks such as the LSTM method used in SHRED. Strikingly, in all examples the accuracy of reconstruction deteriorates only very slightly with increasing depth.

\begin{figure}[ht]
    \centering
    
    \includegraphics[width=\linewidth,]{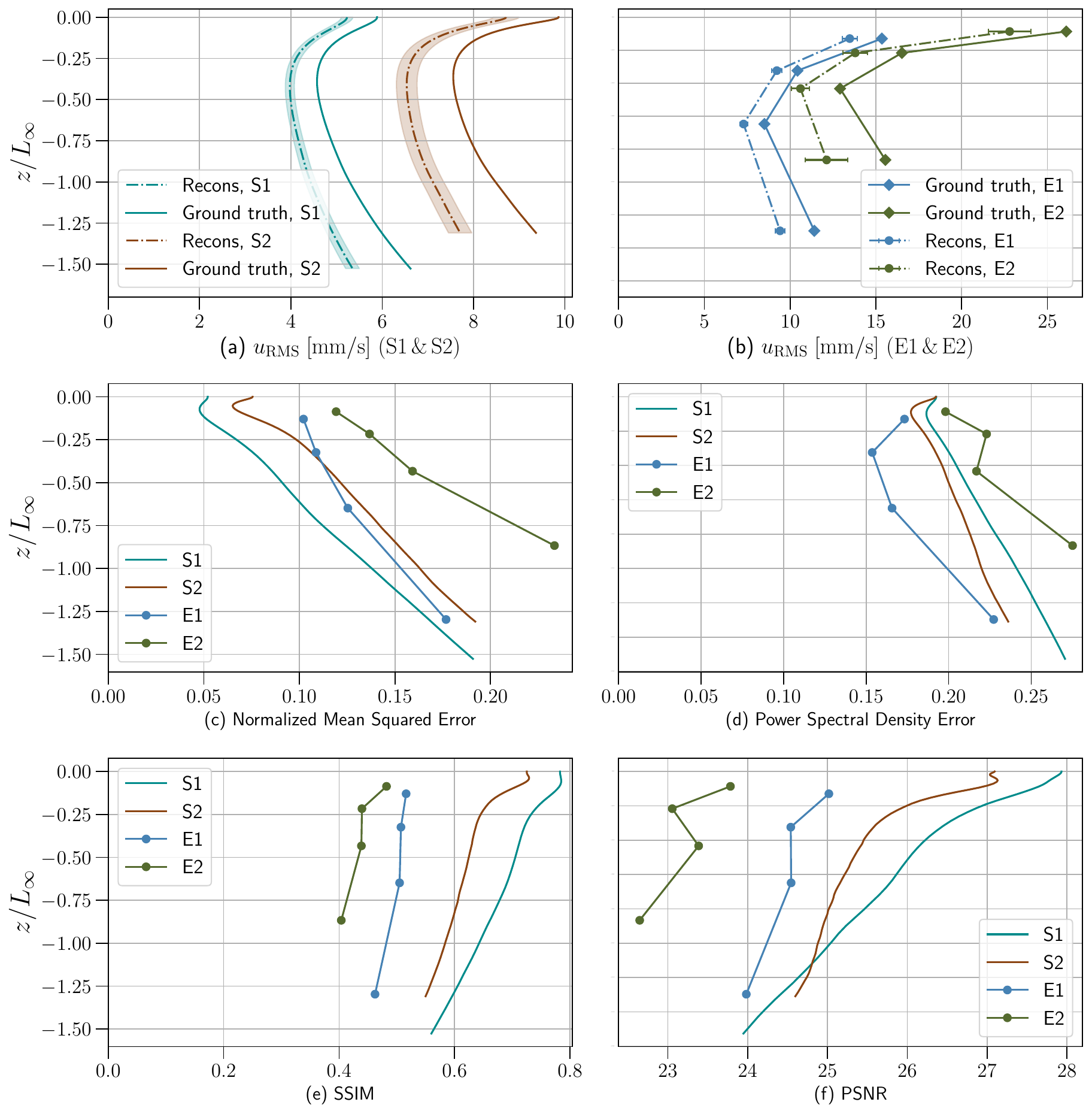}
    \caption{Comparison of error metrics at different depths beneath the surface for the four 
    \rjo{
    cases
    }{datasets}
    in Table \ref{tab:case_details}, taking the ensemble average of $25$ independent SHRED reconstructions. Panels (a) and (b) show reconstructed values of $\langle\urms\rangle$ (dashed lines) for DNS and experimental 
    \rjo{
    cases%
    }{data}
    , respectively, compared with ground truth measurements (solid lines). 
    The standard deviation of the ensembles are shown as shaded areas (a) and error bars (b).
    The bottom four panels show error metrics NMSE (c), PSDE (d), SSIM (e) and PSNR (f), all defined in Sec.\ \ref{sec:method_metrics}. 
\akr{
    All errors are measured relative to the uncompressed fields.
}%
    }
    \label{fig:depth_dep_error}
\end{figure}

Shifting our attention to time-averaged results, the corresponding time-averaged quantities, $\langle\urms(z,t)\rangle$, are plotted in Figs. \ref{fig:depth_dep_error}(a) and \ref{fig:depth_dep_error}(b) for DNS and experimental data, respectively, after averaging over an ensemble of 25 individual SHRED reconstructions each using a different set of three randomly chosen surface points as their sparse input.
By normalizing the depth coordinate by the integral scale $L_\infty$ for each 
\replace{case, }{
flow,
}%
the profiles can be compared side-by-side. 
Again, the solid lines represent the 
\rjo{
original
}{uncompressed}
data (i.e., ground truth), and color-matching dashed lines mark their reconstructed counterparts; the standard deviation of the ensemble is shown as shading [Fig. \ref{fig:depth_dep_error}(a)] or error bars [Fig. \ref{fig:depth_dep_error}(b)]. 
As was observed at individual instants in Fig.\ \ref{fig:rms_instantaneous}, the trends of the profiles are retained by the reconstruction, but consistently offset to lower values, corresponding to a loss of energy in the reconstruction. This can only partly be explained by the compression which truncates the number of SVD modes; the offset is larger than 
expected by a rank reduction alone, as using a moderate truncation value for $r$ has a small impact on the large, energy-carrying scales of the flow (as seen in Fig.\ \ref{fig:SVD_DNS}). 
\ajo{
For an in-depth discussion of the effect of data compression on the results from SHRED, see Appendix \ref{sec:app1}.
}%

The depth-variations of the normalized mean squared error (NMSE) of Eq.\ (\ref{eq:NMSE})
are shown in Fig.\ \ref{fig:depth_dep_error}(c). For the DNS 
\replace{cases,}{
flows
}%
we notice a low MSE of about $6\%-7$\% close to the surface, and a steady increase up to around $19$\% for the lowermost planes. These are generally significantly lower
\ajo{%
values
}%
than the subsurface reconstruction MSE of the CNN method of Xuan \& Shen \cite{xuan2023} 
\akr{%
achieved, although it should be noted that this comparison is not a direct one-to-one benchmark due to differences in configurations. Differences aside, we note that both SHRED and the CNN-based method proposed in \cite{xuan2023}
}%
    \djo{This might suggest that although both methods} 
tend to result in dampened 
\rkr{
amplitudes
}{magnitudes} 
    of large-scale turbulence, 
\rjo{
    with SHRED performing better for this type of simulation data. The higher NMSE
}{for the surface-based flow reconstructions. That the NMSE is higher}
in the lower planes might be expected since there is hardly any correlation in a direct sense between the surface motion and the flow field at this depth.  

 We observe that the 
 \ajo{%
 1D
}%
PSD error metric (PSDE) follows similar trends as the NMSE.
The errors are generally in the range of $15$\%-$30\%$
 in the large-to-intermediate length scales. The reconstructions of the DNS data show similar power spectral
 \djo{energy }%
 errors, although the high-$Re$ 
 \add{%
 data
 }%
 case S2 performs slightly better by this metric. The experimental 
 \replace{case}{
 flow
 }%
 E1 shows lower errors than the DNS cases, which is also reflected in the PSD spectra of Fig.\ \ref{fig:psd_compare}. This is, perhaps, unexpected because the data are generally noisier and harder to reconstruct, as highlighted by the other metrics. 
In contrast, the second experimental 
\replace{case}{
 flow,
 }%
 E2, has the largest PSD error.
 We shall see below that for all 
 \replace{cases, }{
 datasets
 }%
 the PSD error is always associated with a loss in turbulent kinetic energy in the reconstructions, at all lengthscales. Other turbulence-sensing reconstructions find similar losses in turbulent kinetic energy (see \cite{xuan2023, Cuellar2024}), although these are hard to compare directly with our results due to different flow 
 \replace{cases }{%
 conditions
 }%
 and selection of error metrics.

The SSIM results are presented in \ref{fig:depth_dep_error}(e). The SSIM metric differs from the other error metrics that we use, as it targets structure, luminosity, and contrast, while the physical interpretation of the metric is not as obvious.
The SSIM results reflect the qualitative observations made from Figs.\ \ref{fig:reconstruction_RE2500} and \ref{fig:reconstruction Exp}, that there is an indisputable visual similarity between reconstructed fields and ground truth when the larger structures of the turbulent fields are considered.
The DNS reconstructions yield SSIM values between $0.7$ and $0.8$ near the surface, falling to between $0.55$ and $0.65$ \rkr{two}{at one} integral length \rkr{scales below the surface}{scale depth}. For the experimental 
\replace{cases,}{
flows,
}%
peak values are found near the surface at around $0.47$ to $0.52$, falling to around $0.4$ to $0.45$ in the deepest planes. While the latter might be considered a weak result in general image reconstruction schemes, not so in a turbulence context, considering that it is based on the time dynamics of three surface points only, after training. 
The SSIM likely emphasizes small-scale structures of the full-rank ground truth which are no longer present in the compressed training data (or, if they were, would be subject to overfitting). This is particularly prominent for the SSIM results for experimental data.
For the DNS data, the reconstruction SSIM of almost $0.8$ near the surface based on very sparse measurements, can be compared to super-resolution schemes for reconstructing full-state space from a coarse-grained and sparse field. The results of super-resolution reconstructions differ from case to case, but a 
\replace{case study on a }{
study based on data from
}%
turbulent 
\replace{LES}{%
large-eddy simulation (LES)
}%
found many methods to produce values of $0.8$-$0.9$ \cite{Chen2024}. In this regard, SHRED is close to matching this performance 
while also producing acceptable SSIM results for planes far from the surface. 

For the SHRED reconstructions, it is clear that the PSNR values, as seen in panel (f) of Fig.\ \ref{fig:depth_dep_error}, are within acceptable values (above 20 dB) for all 
\replace{cases.}{%
datasets.
}
We are not aware of other sufficiently similar reconstruction studies to which the values can be directly compared, yet one might note that 
super-resolution reconstructions of turbulent DNS data have found similar PSNR values
\cite{pant2021}. The performance is matched, even with planes as deep as $L_{\infty}$ or further from the surface, with PSNR values of 22.5-25.5 dB. The DNS 
\replace{cases show }{
datasets yield
}%
higher values than the experimental ones, whereas the more turbulent ones show lower values. As expected, the PSNR value generally decreases with depth, with the largest decrease occurring near the surface in all cases, where the flow changes rapidly due to surface viscous and blockage effects from the surface \cite{shen99,aarnes2025}.

\begin{table}[!htbp]
\centering
\caption{
\akr{
Energy retained in the reconstructed velocity fields, relative to the energy in the ground truth data. Computation as a cumulative measure of energy retained from the surface down to a fixed depth, according to Eq.~\eqref{eq:rel_energy}.}
}%
\setlength{\tabcolsep}{8pt}
\begin{tabular}{l |c | c | c}
\hline
\textbf{Dataset} & $E(z \leq 0.1\,L_{\infty})\;(\%)$ & $E(z \leq 0.5\,L_{\infty})\;(\%)$ & $E(z \leq 1.0\,L_{\infty})\;(\%)$ \\
\hline
S1 & 79.0   & 78.3    & 76.9   \\
S2 & 78.7  & 77.6  & 75.8  \\
E1 & 75.9   & 75.3    & 73.7   \\
E2 & 72.6   & 71.3  & 69.6   \\
\hline
\end{tabular}
\label{tab:energy_loss_main}
\end{table}

\akr{
The dampened magnitudes in the SHRED reconstructions, which we observe in the the depth-profiles of $\urms$ in Figs.~\ref{fig:rms_instantaneous} and \ref{fig:depth_dep_error}(a)--(b), indicate a systematic energy loss. We quantify how much energy SHRED is able to retain in the reconstructions, by using $ \langle \urms^2(z_k)\rangle $ as a single-component measure of the kinetic energy of a horizontal velocity plane at depth $z_k$. The relative energy retained from the surface to a selected depth $z_l$ is then computed as a cumulative measure by:
\begin{equation} \label{eq:rel_energy}
    E(z \leq z_l) = \frac{\sum_{z_k=z_0}^{z=z_l} \langle \tilde{u}_{\mathrm{RMS}}^2(z_k)\rangle }{ \sum_{z_k=z_0}^{z=z_l} \langle u_{\mathrm{RMS}}^2(z_k)\rangle} \, ,
\end{equation}
where $\tilde u$ and $u$ denote the  reconstruction field and ground truth, as above, and $z_0$ denotes the posistion of the uppermost plane for velocity measurements in each dataset (coinciding with the surface for DNS datasets, but not for experiments).
}

\akr{
The relative energy retained down to three different depths is reported in Table \ref{tab:energy_loss_main}. Note that as the experimental datasets E1 and E2 are sparse in depth, calculation down to $z_l = 0.1 L_{\infty}$ includes only the top plane, $z_l = 0.5 L_{\infty}$ includes the three uppermost planes, and $z_l = 1.0 L_{\infty}$ includes all four planes.
We observe that the retained energy decreases gradually with depth for all datasets, consistent with the increase of errors and reduction of surface-to-subsurface correlation with depth, with roughly $70$\%-$80\%$ of the energy captured by the SHRED reconstructions down to one integral length scale below the surface. Although one might expect that much of the energy loss is due to the fact that SHRED exclusively trains and validates on a significantly compressed dataset, the compression itself only accounts for 2-7 percentage points of the energy loss (details in Appendix \ref{sec:app1}).
A glance at Fig.~\ref{fig:psd_compare}, which we consider in detail later, indicates that it is SHRED's energy loss at the large scales (low wavenumbers) which is the major driver of the total energy loss. The effects noted here may be tied to the problem of properly representing the large, highly intermittent upwelling events in our free-surface flow. The very largest of these are rare, e.g., occuring only 5--10 times in each of the DNS datasets, making it very hard to fully account for them across our training data. The LSTM, despite good temporal performance, is best suited for learning smooth dynamics and together with the decoder, they might underestimate intermittent magnitudes of energetic scales.
}

Overall, the depth-dependent error metrics indicate that even \rkr{$2-2.5$}{1.0--1.5}integral length scales from the surface, large-to-intermediate-scale turbulence can be reconstructed well enough for many practical purposes from just three measurement points of the surface elevation, within the time range of training data. As a proof-of-concept study these results demonstrate the potential of SHRED for remote sensing applications, where only 
observations \emph{at} the surface, and not beneath it, 
are available. The ability to reconstruct bulk flow structures using just three surface points demonstrates a key step toward remote sensing of subsurface turbulence. However, we emphasize that although the timesteps where reconstruction is performed are not part of the actual training set, they lie within the same time range used in training.
Future work will explore the capability of SHRED to make reconstructions of previously unseen flow regimes or truly independent test cases, which is crucial for \ajo{generalizability and 
}%
real-world deployment.

\subsection{\ajo{Comparison with linear, POD-based method}}

\begin{table}
\centering
\caption{
\akr{
Performance comparison of SHRED and POD-based linear regression at multiple depths for datasets S2 and E2.
}
}
\label{tab:pod_vs_shred}
\begin{tabular}{c c c |c| c c c}
\hline

& & & \multicolumn{1}{c|}{\textbf{SHRED}} & \multicolumn{3}{c}{\textbf{POD + linear regression}} \\
\cline{4-7}
\textbf{Dataset} & \textbf{Depth} & \textbf{Metric} 
& 3 sensors 
& 3 sensors & 30 sensors & 300 sensors \\
\hline

\multirow{6}{*}{S2}
& $z/L_{\infty}\approx 0.1$ & NMSE   & 0.072 & 1.028 & 1.861 & 0.029 \\
&                           & SSIM   & 0.704 & 0.043 & 0.124 & 0.775 \\
\cline{2-7}

& $z/L_{\infty}\approx 0.5$ & NMSE   & 0.123 & 1.028 & 1.749 & 0.070 \\
&                           & SSIM   & 0.625 & 0.042 & 0.100 &0.688 \\
\cline{2-7}

& $z/L_{\infty}\approx 1.0$ & NMSE   & 0.161 & 1.020 & 1.786 & 0.451 \\
&                           & SSIM   & 0.582 & 0.048 & 0.105 & 0.593 \\
\hline

\multirow{6}{*}{E2}
& $z/L_{\infty}\approx 0.1$ & NMSE   & 0.124 & 1.527 & 0.584 & 0.706 \\
&                           & SSIM   & 0.486 & 0.048 & 0.316 & 0.405 \\
\cline{2-7}

& $z/L_{\infty}\approx 0.5$ & NMSE   & 0.158 & 1.276 & 0.284 & 0.309 \\
&                           & SSIM   & 0.432 & 0.012 & 0.356 & 0.335 \\
\cline{2-7}

& $z/L_{\infty}\approx 1.0$ & NMSE   & 0.240 & 1.252 & 0.440 & 0.422 \\
&                           & SSIM   & 0.402 & 0.031 & 0.267 & 0.296 \\
\hline

\end{tabular}
\end{table}
\akr{To evaluate SHRED as an overall sensing and reconstructing method, we compare its performance with that of proper orthogonal decomposition (POD) of fields with linear regression estimation for sensors-to-POD mapping. The linear regression estimation uses least-squares regression mapping from lagged sensor measurements to $r$ number of POD coefficients, where we set $r$ equal to the low-rank values from Table \ref{tab:SVD_rank}.} \ajo{We perform the comparison on the the highest Reynolds number dataset from the DNS and experiments, datasets S2 and E2, respectively, and compute the NMSE and SSIM for multiple depths for each dataset. As in \S \ref{sec:depth_error_results}, we report ensamble-averaged error metrics, with 25 ensemble runs used for the SHRED computations and 10 for the POD with linear reconstruction. The results are presented in Table \ref{tab:pod_vs_shred}.}

\akr{
While POD provides an efficient low-dimensional representation of the flow, the results in Table \ref{tab:pod_vs_shred} indicate that recovering its coefficients from linear mapping of sparse surface measurements is highly challenging. With the identical three sensor set-up that we have so far used for SHRED reconstructions, 
the POD-based linear regression mapping performed poorly, with NMSE above 1 and SSIM close to zero, indicating that 
errors exceeds signal energy and that spatial structures are not well preserved in the output. In short, for the identical sparse-sensor configuration as SHRED, the POD-based linear regression method comes nowhere near SHRED's performance.
}

\akr{
Additional tests show that POD sensing performance depends strongly on the number of surface sensors, whereas the dependence on retained rank is weak. We therefore include POD sensing results for configurations with 30 and 300 sensors to level the playing field between the two methods we compare here. As the POD-based regression method is linear, performance generally improves with more inputs. SHRED, on the other hand, is nonlinear, and its performance improves with sensor count only to a certain point, while high input dimensions increase complexity of the learning and degrades reconstructions (details in Appendix \ref{sec:app2}). From the POD-regression analysis, we find that approximately two orders of magnitude more input sensors are needed to match the ultra-sparse 3-sensor SHRED performance for our datasets. However, even with more sensors, the magnitude of deeper velocity planes (here: $z/L_{\infty} \approx 1$), are poorly captured by the POD-based linear regression. Hence, although the POD-based linear method yields comparable results to SHRED for dense inputs, SHRED outperforms it substantially in the sparse-sensor regimes, as well as for surface-to-depth sensing, and is thus more powerful for turbulence sensing of limited data. These findings are consistent with previous studies demonstrating nonlinear decoder networks outperforming linear methods in sparse regimes \cite{Erichson2020}, as well as the findings by Williams et al.~\citep{williams2024}, showing SHRED outperforms linear POD-based methods for a range of high-dimensional spatiotemporal datasets, including isotropic turbulence.
}

\FloatBarrier

\subsection{Temporal analysis of planar root-mean-square velocity}
\begin{figure}[!htbp]
    \centering
    \begin{overpic}[width=\textwidth, trim={0cm 0cm 0cm 0cm}, clip, grid=off, unit=2bp,tics=2]{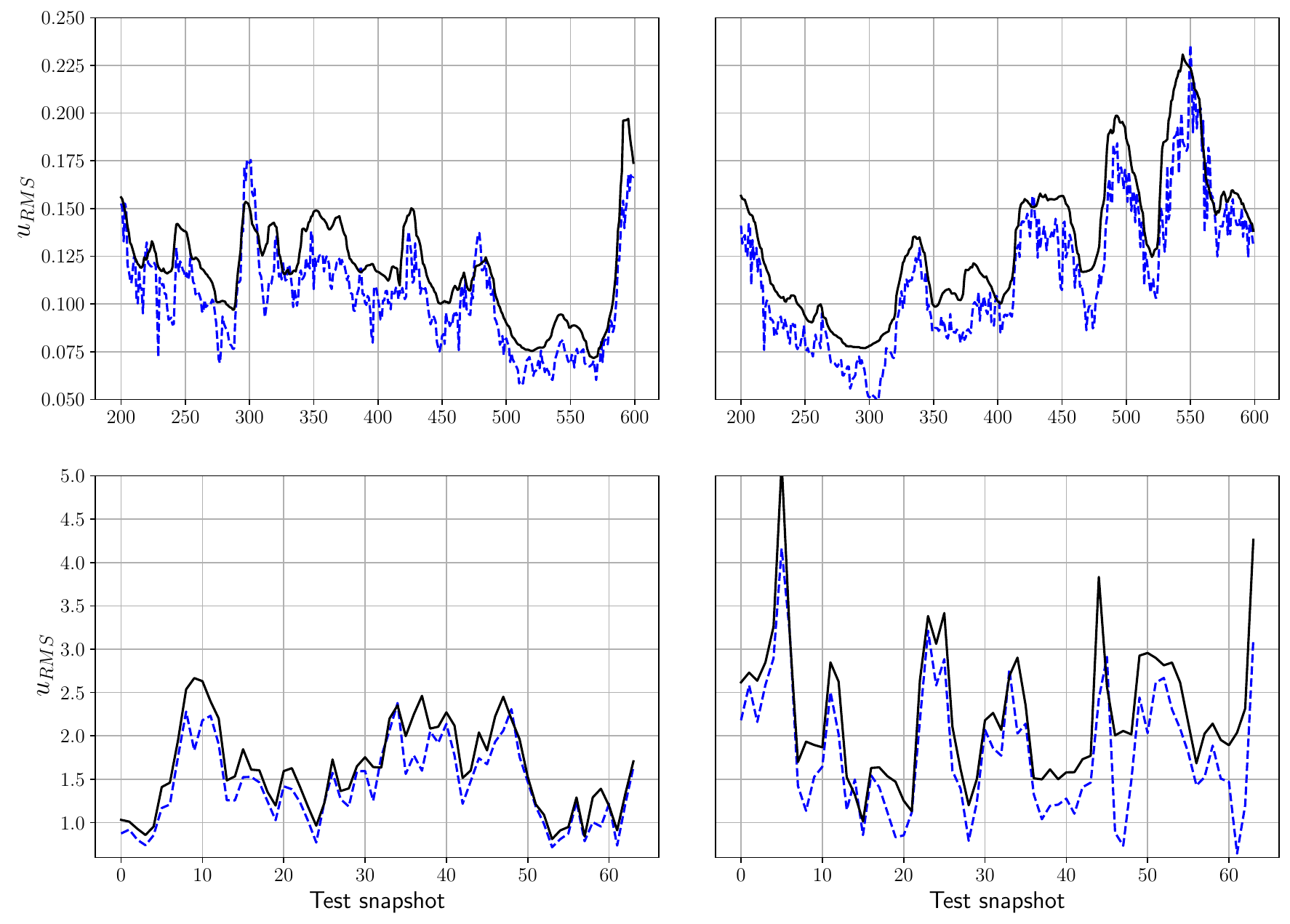}
    \put (15,68) {\makebox[0pt]{\Centerstack{S1}}}
    \put (65,68) {\makebox[0pt]{\Centerstack{S2}}}
    \put (15,32) {\makebox[0pt]{\Centerstack{E1}}}
    \put (65,32) {\makebox[0pt]{\Centerstack{E2}}}
    \end{overpic}
    \caption{Examples of ground truth (black solid line) and reconstruction (blue dashed line) of single-frame rms velocity, $\urms$, a time series of test snapshots near the surface
    \add{
    for all four datasets in Table \ref{tab:case_details}
    }
    at depths of around \rkr{$0.14-0.20$}{$0.08-0.12$}$L_{\infty}$.}
    \label{fig:rms_time_series}
\end{figure}
As an illustration of the SHRED reconstruction capabilities of temporal dynamics, we show the time series of the planar RMS values for a single SHRED reconstruction case, in Fig.\ \ref{fig:rms_time_series}. The planes chosen are relatively close to the surface, 1~cm depth for the experimental 
\replace{cases }{
flows
}%
($0.21 L_{\infty}$ for E1 and $0.14\,L_{\infty}$ for E2), and planes at $0.2\,L_{\infty}$ depths for DNS 
\add{
flow
}%
cases S1 and S2. 
We observe a 
\rjo{%
close
}{high degree of}
correlation between the time series of the ground truth and the reconstruction. The normalized cross-correlation values with zero lag, are above $0.94$ for all 
\replace{cases except case }{
datasets except
}%
E2, in which the correlation value is $0.89$. The values of $\urms$ are generally lower for the reconstructed fields than the ground truth, indicating some loss of 
\djo{
turbulent
}%
kinetic energy%
\ajo{%
, as discussed in \S \ref{sec:depth_error_results}%
}%
. Notably, the error in $\urms$ does not 
\rjo{%
noticeably increase
}{increase dramatically}
when intermittent high-intensity turbulent events,
\djo{ occur,} 
corresponding to sudden peaks in the RMS velocity%
\ajo{%
, occur.
}

\subsection{Spectral analysis}
\label{sec:spectra}
\begin{figure}[!htbp]
    \centering
    
    \begin{overpic}[width=\textwidth, trim={0cm 0cm 0cm 0cm}, clip, grid=off, unit=2bp,tics=2]{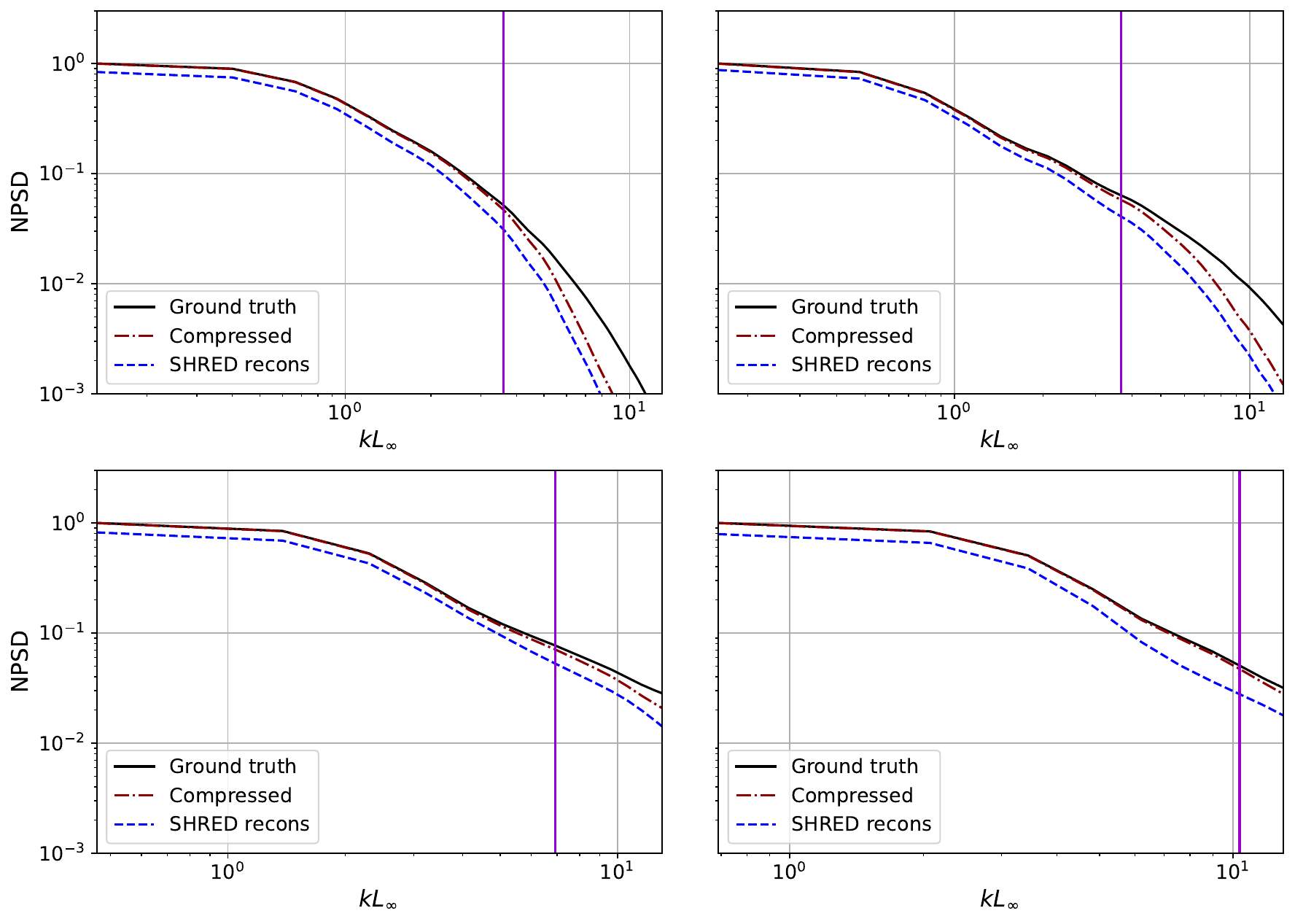}
    \put (15,68) {\makebox[0pt]{\Centerstack{S1}}}
    \put (65,68) {\makebox[0pt]{\Centerstack{S2}}}
    \put (15,32) {\makebox[0pt]{\Centerstack{E1}}}
    \put (65,32) {\makebox[0pt]{\Centerstack{E2}}}
    \end{overpic}

    \caption{\rjo{Power density spectra}{Normalized 1D power spectral density (NPSD)} for the four 
    \replace{cases, }{
    datasets,
    }%
    \rjo{showing full-rank
    }{including results for the uncompressed, }
    ground truth data (dark, solid line), low-rank 
    \rjo{SVD truncation
    }{compressed data}
    (dash-dotted, red line) and SHRED reconstructions 
    \djo{spectra }%
    (blue, dashed line), for the velocity fields at depths $\approx 1.0 L_{\infty}$. The purple vertical line indicate the cutoff wavenumber for the SVD
    \ajo{%
    compression
    }%
    , i.e., the highest resolved wavenumber in the 
    \rjo{
    SVD
    }{velocity fields of the compressed data}
    and SHRED reconstructions.
    \djo{ velocity fields.}
    The spectra are normalized by the maximal value of the ground truth spectra, and the wavenumber is normalized by the inverse integral length scales in each case. }
    \label{fig:psd_compare}
\end{figure}

In this section, we investigate how the power density spectra of SHRED-reconstructed fields perform compared to \ajo{uncompressed data
}%
(ground truth) and the
\ajo{ %
compressed,
}%
low-rank 
\rjo{SVD truncation.}{
data.
}%
An example is shown in Fig.\ \ref{fig:psd_compare}, for 
\rjo{a SHRED case}{
reconstruction of flow
}%
at a depth of 
\rkr{$1.5 L_{\infty}-2.0 L_{\infty}$.}{
approximately $ L_{\infty}$.
}%
The 1-D power density spectrum in the $x$-direction, averaged over time and spatially in the $y$ direction, is calculated 
\replace{for}{
in
}
each case. The purple dotted vertical line indicates the cutoff wavenumber for the low-rank truncation, i.e., the wavenumber at which the compressed spectrum deviates from ground truth by about $10$\%. (See Section \ref{sec:compression} and Appendix \ref{sec:app1} for details on selecting the rank truncation). One observes that generally, the spectra for the compressed field (dashed dotted red line) closely follow the spectra of the ground truth (solid dark line) up to the cutoff wavenumber. We observe that the spectra of the SHRED reconstructions generally follow the ground truth, but with a spectral energy loss 
\rkr{evenly distributed over all frequencies}{
that is fairly consistent in log-space over all frequencies.
However, there are somewhat larger relative errors in the high-frequency end, suggesting fine-scale dynamics are more challenging to capture. On the other hand, the majority of the total kinetic energy is contained in the lower wavenumbers, hence the low-frequency spectral errors contribute the most to the energy loss.
}%

\FloatBarrier
\section{Conclusions}
\label{sec:conclusion}

In this study, we demonstrated a proof-of-concept application of the SHallow REcurrent Decoder (SHRED) for reconstructing subsurface turbulence fields from sparse surface measurements in free-surface flows. Using only three arbitrarily placed sensors measuring surface elevation as input, SHRED was able to infer the dominant large- and intermediate-scale structures below the surface across four different 
\replace{turbulent cases, }{
datasets of turbulent flow beneath a free surface together with the instantaneous elevation of the surface itself,
} two based on simulation data and two on experimental data
\add{
recently published in \cite{aarnes2025} and \cite{babiker2025}.
}%
The differences between the datasets in terms of Reynolds number, sparsity, and noise demonstrate the flexibility of SHRED in handling different turbulence regimes.

The results show that SHRED preserves 
\rkr{important flow features such as spatial structure and spectral content,}{
the most energetically significant flow features,
}%
and performs particularly well near the surface, even 
\rjo{in noisy}{
on highly multi-scale
}%
experimental data. While reconstruction accuracy decreases with depth, SHRED still provides meaningful results as deep as 
\rkr{two integral length scales}{
one integral length scale
}%
below the surface. Importantly, this was achieved without 
\ajo{
direct
}%
access to any subsurface measurements at inference time, highlighting the potential of SHRED as a tool for remote sensing of subsurface turbulence
\ajo{%
from very sparse surface measurements. Note, however, that subsurface fields are necessary during the training step, limiting SHRED to datasets which have (at least partially) temporally resolved flow fields below the surface.
} 

This work addresses the central challenge of estimating near-surface turbulence in rivers and oceans from surface observations alone. Such capability is crucial for quantifying gas and heat fluxes at the water-air interface, where in-situ measurements are impractical at scale. The demonstrated ability of SHRED to learn nonlinear mappings from sparse input to high-dimensional turbulent states marks a step forward toward scalable, non-intrusive field sensing.

Future work should aim to improve generalization across flow regimes, as the current  validation setup is limited to data drawn from the same underlying datasets. 
\ajo{
Moreover, recent research \cite{kutz2024shallow} has demonstrated that SHRED has potential for forecasting in time, and demonstrating this for the complex case of free-surface turbulence would be most advantageous. Due to the intermittency of the largest structures we observe in our datasets a longer time series than we have available is necessary to achieve full flow field forecasting, and we have limited the present study to spatial reconstruction and temporal inbetweening.
}%
Pairing SHRED with other methods, like 
\rjo{SINDy for sparse}{
in SINDy-SHRED \cite{gao2026}, which combines sparse sensing and reconstruction with
}%
identification of nonlinear dynamics%
\ajo{%
, or SENDAI \cite{zhang2026sendaiARXIV}, which includes learned residual corrections across spatial scales,
}%
could potentially improve the reconstruction accuracy outside of the training time domain and could help make the model generalizable to real-world flows. Other possible steps towards remote sensing could be to extend the model to reconstruct derived quantities such as energy fluxes or gas exchange rates. Ultimately, SHRED offers a foundation for machine learning-based frameworks for remote sensing, and opens a path toward real-world applications in oceanography, river monitoring, and environmental sensing.

\section*{Code and data availability}
All code used in producing these results, are included and thoroughly presented in \cite{Moen2025_SHRED_repo}. 
The DNS data and supporting codes are available from \cite{aarnes2025b}%
\add{%
, the experimental data are available from \cite{babiker2026}.
}
\delete{Due to size constraints, raw PIV data are not hosted in the repo; download instructions are included in the GitHub
README. }

\subsection*{Acknowledgements}

\add{
The experimental datasets are from an experiment performed by Ali Semati, Dr.\ Am\'{e}lie Ferran and Dr.\ Yi Hui Tee under the guidance of Prof.\ R.~Jason Hearst; beyond the experimental data itself we have benefited from discussions with these and input on the manuscript. 
}%
DNS data was generously shared with us by Prof.\ Lian Shen and Dr.\ Anxing Xuan at the University of Minnesota. 
\add{
We thank Omer M.~Babiker for assistance with data handling and preparation, and many discussions.
}%
The work of
\delete{KSM, }%
JRAa and S\AA E was co-funded by the Research Council of Norway (\emph{iMOD}, grant 325114) and the European Union (ERC CoG, \emph{WaTurSheD}, grant 101045299). 
Views and opinions expressed are however those of the authors only and do not necessarily reflect those of the European Union or the European Research Council. Neither the European Union nor the granting authority can be held responsible for them.
The work of JNK was supported in part by the US National Science Foundation (NSF) AI Institute for Dynamical Systems (dynamicsai.org), grant 2112085.   JNK further acknowledges support from the Air Force Office of Scientific Research  (FA9550-24-1-0141).

\add{
\subsection*{Author contributions}

KSM and JRAa performed the data analysis with SHRED and wrote the first draft of the majority of the manuscript; they are to be considered joint first authors. JNK made the first implementation, co-developed the concept and contributed writing and supervision. S\AA E contributed to the concept and discussions, supervision. All authors contributed significantly to the writing of the final manuscript.
}


\appendix
\section{\rkr{Parametric study of optimal rank value}{Effect of SVD compression on SHRED performance}}
\label{sec:app1}

\ajo{
As discussed earlier, we compress the datasets before training and validation of SHRED. The compression is done by truncating the SVD representation of the flow to a fixed rank $r$, hence, the compressed dataset contains only the $r$ most energetic SVD modes. The choice of rank has a significant impact on the reconstruction performance of SHRED.
}%
\djo{When compressing the input velocity field and surface height, only the $r$ most energetic modes of the SVD are used. In order to study the performance of SHRED in simulation and experimental data, it is of interest to first find a suitable rank value $r$ for the SVD. }
If
\ajo{
the value of $r$
}%
is too low, relevant small-scale turbulence might be lost. If too high, too much noise might be included, with a risk of overfitting.
\ajo{
In the present section, we include details on the selection of rank for all four datasets and investigate the effect of data compression on the depth-dependent error metrics and energy retention which we considered in \S \ref{sec:depth_error_results}.
}%
\djo{One way to find an optimal rank $r$ is to test a range of different values, run SHRED, and compare the reconstructed fields with the original uncompressed fields. The best value would be that which results in a reconstruction as close as possible to the ground truth, as measured by the chosen error metrics.}


\subsection{\akr{Parametric study of optimal choice of rank}}

\begin{figure}[ht]
    \centering
    \includegraphics[width=\linewidth]{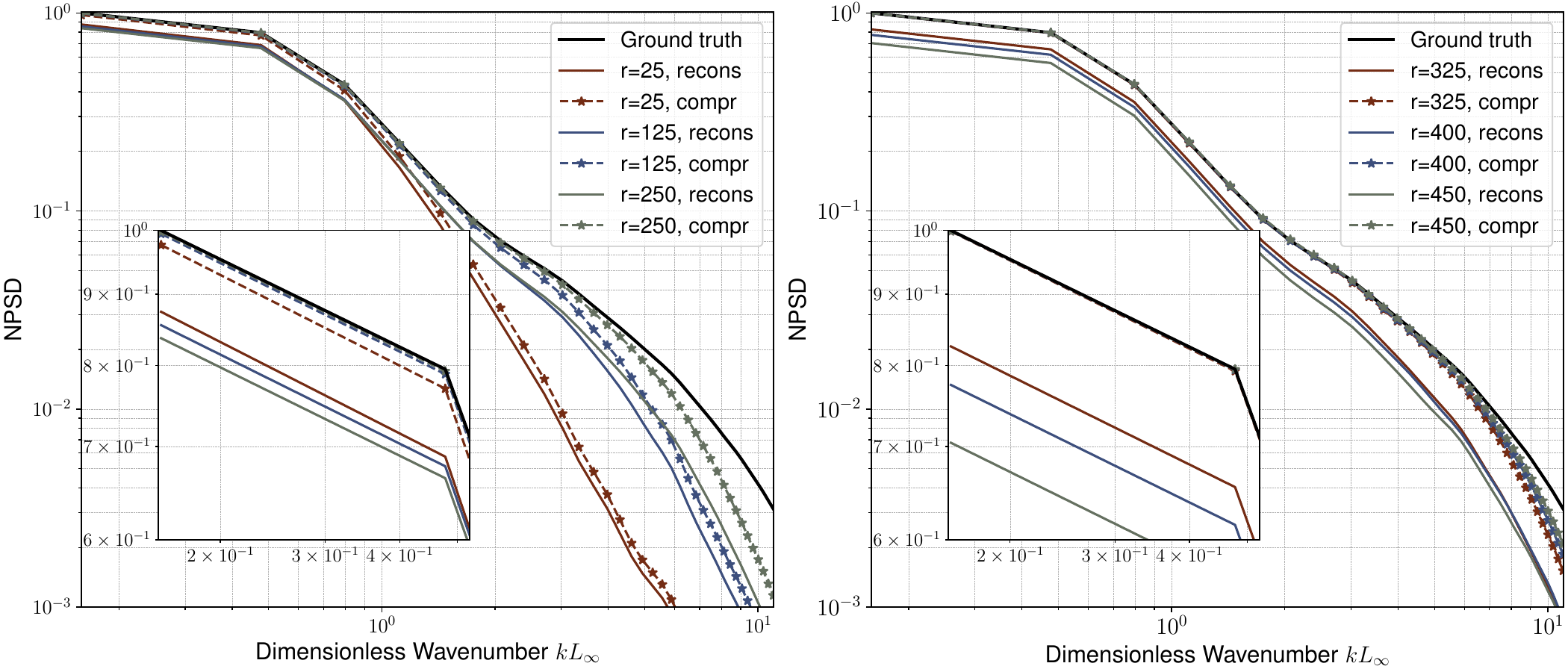}
    \caption{\rjo{Turbulent spectra from data compressed with}{Normalized 1D power spectral density (NPSD) of the}
    low- and moderate-rank 
    \rjo{
    data
    }{value $r$}
    (left) and high-rank 
    \rjo{
    data
    }{value}
    (right) for 
    \replace{case }{
    dataset
    }%
    S2. Solid lines represent data reconstructed by the SHRED model, dashed-stared lines compressed data and the solid black line the ground truth spectrum. Inset: Zoomed in view of the largest scales.}
    \label{fig:spectra}
\end{figure}

\rjo{
To get a first impression of the effect of compression on the flow and the reconstruction with SHRED, consider Fig.\ \ref{fig:spectra}. We
}{From Fig.\ 12 we}
observe that the \ajo{1D} PSD spectra for the SHRED-reconstructed fields (full colored lines) generally deviates from both the compressed (dashed-star lines) and the ground truth (black solid line) for a range of 
\rjo{
chosen rank.
}{compression levels.}
Specifically, in the energy-containing region at low wavenumbers, the compressed spectra follow ground truth closely
\ajo{ for all but the very highest compression level (lowest rank, $r=25$)%
}%
, whereas the reconstructed field spectra appear to
\rjo{
be more reduced with
}{deviate further from ground truth for the}
higher rank 
\rjo{
truncation.
}{values (right panel),}
We take this as an indication that it is SHRED itself, not the SVD compression, that contributes the most to the spectral error of reconstructed fields for the large-to-intermediate scales. \ajo{We return to this point in \S \ref{sec:app1}2.}

\begin{figure}[!htbp]
    \centering
    
    \begin{subfigure}[b]{0.48\textwidth}
         \centering
         \includegraphics[width=\textwidth]{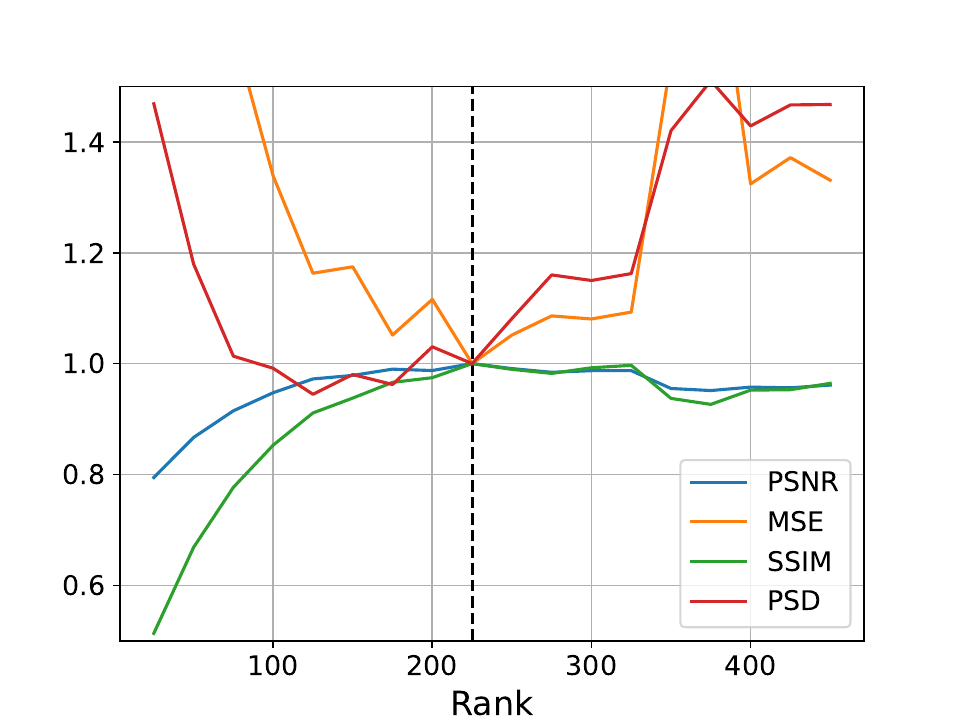}
         \caption{S1}
         
     \end{subfigure}
              \begin{subfigure}[b]{0.48\textwidth}
         \centering
         \includegraphics[width=\textwidth]{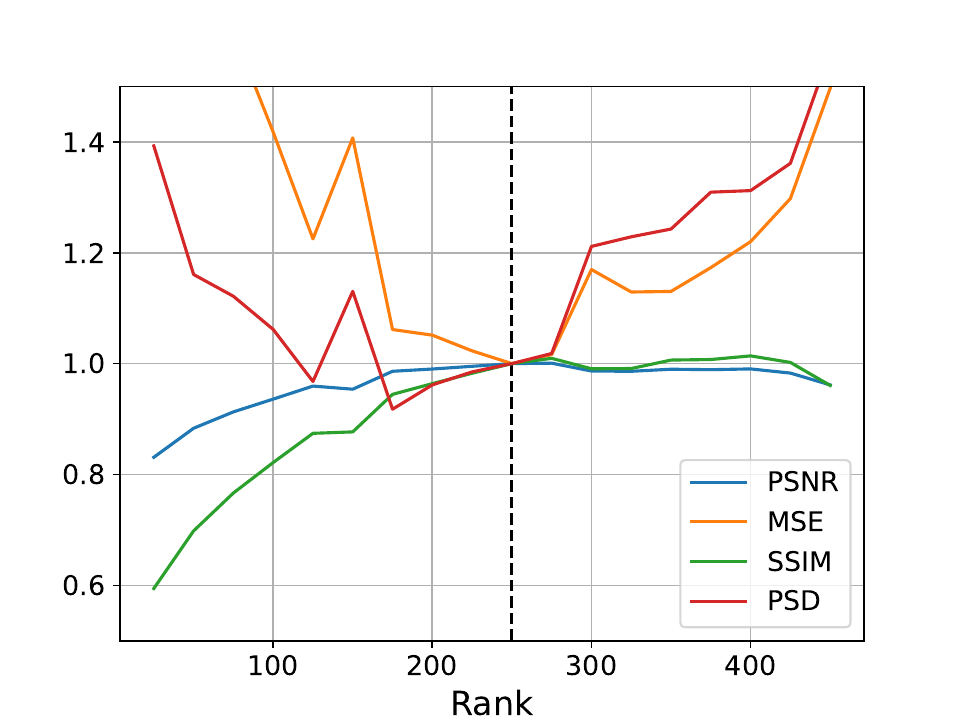}
         \caption{S2}
        
     \end{subfigure}

    \vspace{0.3cm}
     
     \hfill
     \begin{subfigure}[b]{0.48\textwidth}
         \centering
         \includegraphics[width=\textwidth]{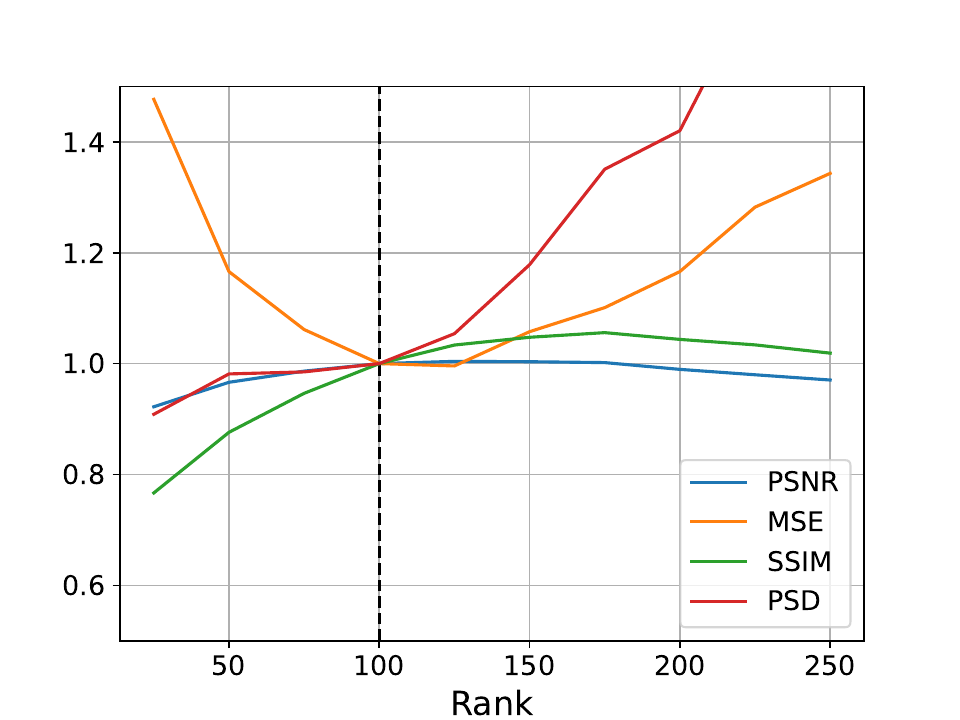}
         \caption{E1}
         
     \end{subfigure}
          \begin{subfigure}[b]{0.48\textwidth}
         \centering
         \includegraphics[width=\textwidth]{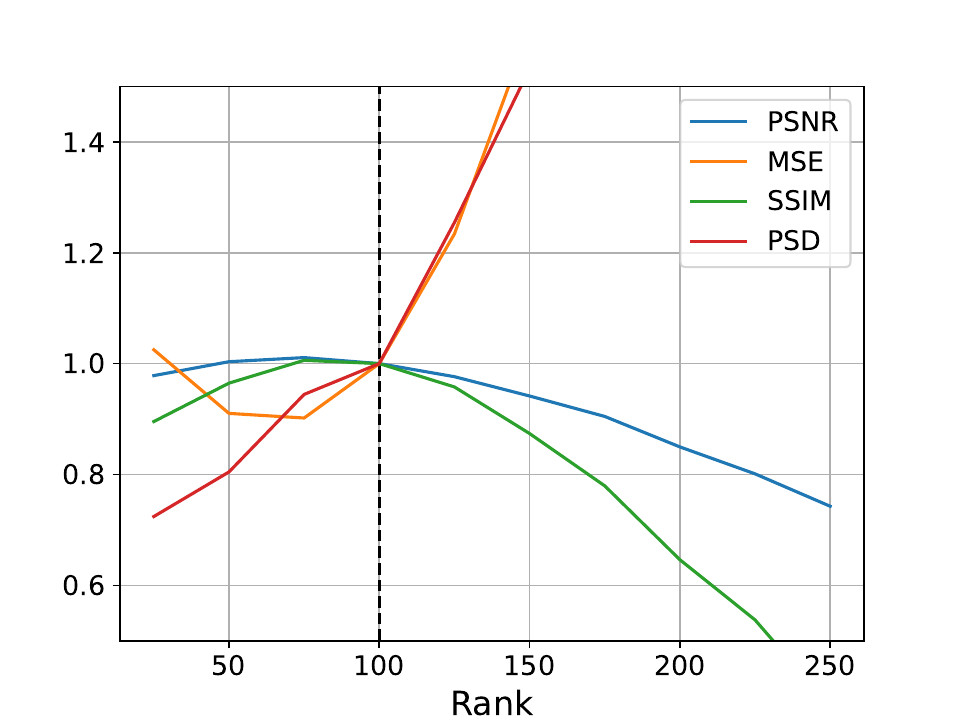}
         \caption{E2}
         
     \end{subfigure}
     \hfill
    \caption{Normalized error metrics averaged in depth for different 
    \rjo{cases of SVD ranks.}{
    levels of data compression, where higher rank denotes less compression.
    } 
    The dashed line marks the 
    \rjo{chosen values for the DNS and experimental cases. In the former case, $r=250$ is chosen as the optimal SVD truncation for best SHRED performance.}{
    rank used throughout the study (see Table \ref{tab:SVD_rank}).
    } The error metrics are normalized 
    \rkr{
    based on maximal value for better comparison.
    }{relative to the mean error at this rank for each dataset.} 
    Note that optimal SHRED performance is met if PSNR and SSIM are maximal, while the same is true when PSD and MSE are minimal.}
    \label{fig:r_params}
\end{figure}

\label{sec:parameter_rank}
To quantify \rjo{error}{the performance of SHRED} as a function of rank \rjo{number, we make use of the error metrics from \S \ref{sec:method_metrics}, averaging them in depth for SHRED applied to data compressed with}{used in the data compression step, the selected error metrics are averaged depthwise, and calculated for} 
a range of rank \rjo{values.}{truncation values.}
Figure \ref{fig:r_params} shows the rank-dependency of the SHRED error metrics for all four 
\rjo{cases}{datasets. The metrics are normalized by the error values at the rank we use for our main results, i.e. $r=225$ for S1, $r=250$ for S2 and $r=100$ for E1 and E2} for scaling and comparison purposes. 
\rjo{The chosen rank truncation number for all of the SHRED analysis in Sec.\ \ref{sec:results}, are}{This rank value is} marked by the dashed black lines.\djo{ in each panel.} The range of the rank 
\rjo{analysis}{values we test} is different between simulation and experimental \replace{cases, }{
datasets,
}%
due to the difference in total number of modes
\djo{(equal to number of time steps). For example,}%
\ajo{
---
}
the full rank of the experiments is $r=900$. while for the DNS 
\djo{
cases
}%
it is $10900$ for S1 and $12500$ for S2. 

The error metrics indicate that the PSNR and SSIM values increase up until a certain rank number, while PSD and MSE generally decreases to a minimal error \ajo{before increasing rapidly if $r$ is further increased}. This indicates that for too low rank numbers, there is a significant lack of information left in the low-rank representation, such that although SHRED might perform better in reconstructing the compressed input fields, the SVD itself has left out a significant amount of information. Hence, the fields lack structure and contrast for high SSIM values, and amplitudes and energy content of the largest modes are low, making NMSE and PSDE high. On the other hand, the error metrics generally show that SHRED performance decreases if the rank number is too high. In this case, the low-rank representation includes more fine details, including some noise. The temporal dynamics of the structures on these scales is highly intermittent and random, hence notably harder to reconstruct, especially from surface dynamics only. 
\djo{This decrease of performance for higher rank numbers, is caused by SHRED, not by the SVD truncation. }
In between these two regimes, there is a range of rank numbers where the error metrics are minimal. This sweet spot is where one finds optimal rank truncations that balances between having enough SVD modes included, while not over-saturating SHRED with noise and unpredictable small-scale turbulence.

\subsection{\akr{Depth-dependent error error from data compression}}

\begin{figure}[htb!]
    \centering
    
    \includegraphics[width=\linewidth,]{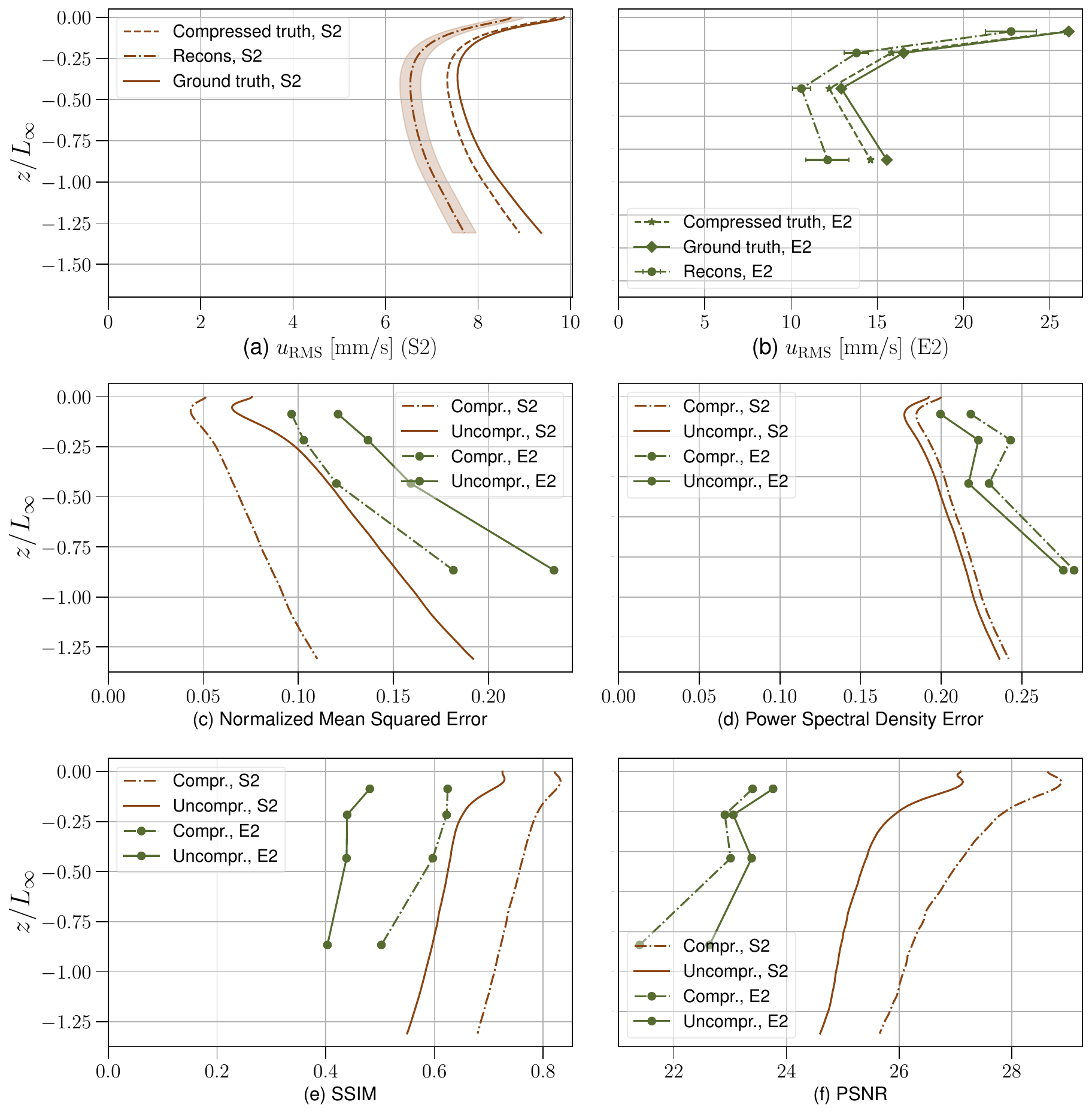}
    \caption{\akr{Same comparison of error metrics as in Fig. \ref{fig:depth_dep_error}, but comparing SHRED performance relative to the true value of the compressed data (dash-dotted curves) and to the uncompressed data (solid lines) for datasets S2 and E2. Panels (a) and (b) show reconstructed values of $\langle\urms\rangle$ for DNS and experimental datasets respectively, with true values for the compressed data added as dashed lines. The standard deviation of the ensembles are shown as shaded areas (a) and error bars (b).
    The bottom four panels show error metrics NMSE (c), PSDE (d), SSIM (e) and PSNR (f), all defined in section \ref{sec:method_metrics}}.
    }
    \label{fig:depth_dep_error_compressed}
\end{figure}
\add{
To gain further insight into the effect of the compression, we isolates the error introduced by compression from that introduced by SHRED in two ways: First, we compare retained energy in the reconstructions over approximately one integral length scale in depth, with respect to the uncompressed data and the compressed data (computation details are reported in \S \ref{sec:depth_error_results}). Secondly, we compute depth-dependent error metrics with the compressed data taken as ground truth and compare them to the error metrics with the full, uncompressed datasets as ground truth from \S \ref{sec:depth_error_results}. Note that for the results presented here, we use the $r$ values listed in Table \ref{tab:SVD_rank} and represented by dashed lines in Fig.~\ref{fig:r_params}.

The retained energy in the reconstructed velocity field is reported in Table \ref{tab:energy_loss}, as a cumulative measure down to  $z \approx L_{\infty}$, relative  to the uncompressed and compressed fields. Measured against the compressed dataset, SHRED to retains roughly $80 \%$ of the ``kinetic energy" for data sets S1, S2 and E1, and 72\% for data set E2. 
When measured against the uncompressed data, the recovered cummulative kinetic energy is 2--7 percentage points lower. In short, SHRED underestimates the kinetic energy, with the main contribution coming from the reconstruction algorithm itself, not from the data compression step.
}

\begin{table}
\centering
\caption{
\akr{
Retained energy measure of the reconstructed fields, relative to the uncompressed and compressed data, respectively. All numerical values in \%.
}
}
\setlength{\tabcolsep}{6pt}
\begin{tabular}{l | cc|}
\hline
& \multicolumn{2}{c|}{$z \leq 1.0\,L_{\infty}$} \\
\textbf{Dataset}

& Uncompressed &  Compressed \\
\hline
S1 & 76.9 & 79.8   \\
S2 & 75.8 & 80.7 \\
E1 & 73.7 & 80.7   \\
E2  & 69.6 & 71.8   \\
\hline
\end{tabular}
\label{tab:energy_loss}
\end{table}

\akr{
We next consider the depth-dependent metrics which compare errors introduced by SHRED and by the data compression, depicted in Fig.~\ref{fig:depth_dep_error_compressed}. In panels (a) and (b), we observe a results which basically reiterates the kinetic energy consideration above: The time averaged $\urms$ from the reconstructed fields is closer to the corresponding curve from the compressed data than that from the uncompressed data, but deviates more from both of these than they do from each other. In panels (c)--(e), the general trend is, as expected, that the metrics are for datasets S2 and E2 are shifted to lower errors and higher SSIM/PSNR, when measured against the compressed fields, rather than directly against the uncompressed data. 
Among these metrics, the NMSE and the SSIM stands out, in showcasing how much better performance we get when measuring against the compressed data (which is the data we actually train and validate on), rather than against the uncompressed data. Measured against the compressed data,  the NMSE, the NMSE is reduced by roughly $0.04$-$0.07$, equivalent to a $20$\%-$40\%$ reduction from the NMSE computed agsinst the uncompressed data. A similar adjustment is seen for the SSIM, which is roughly $0.15$-$0.20$ higher for the comparison with the compressed data. The difference can be attributed to the removal of small-scale structures by the compression, as seen in the field comparisons of Figs. \ref{fig:reconstruction_RE2500} and \ref{fig:reconstruction Exp}, which is captured well by the NMSE and SSIM measures. 
The only measures that deviates from the trend, is the PSNR for dataset E2, where the reconstruction performance is deemed to be slightly better when measured against the uncompressed rather than the compressed data.
}

\section{\ajo{Influence of sensor count on SHRED performance}}
\label{sec:app2}

\akr{
For the main analysis of SHRED, we used an ultra-sparse sensor scheme of three randomly placed surface sensors. SHRED has previously been shown remarkable reconstruction performance across different datasets with this sensor scheme using three sensors \cite{williams2024, Faraji2025, gao2026, tomasetto2025reducedPub}. While this setup showcases the major strength of SHRED compared to other methods, e.g. POD-based regression, or even CNNs, the available sensor data might in some cases be dense, as in multiple sensor points or even high-resolution images. However, as SHRED is a nonlinear network, the effect of having more input data does not necessarily entail improved performance. In the paper where SHRED was first proposed, Williams et al.~\citep{williams2024} analyzed reconstruction dependency on sensor count and placement, using up to 50 sensors. They found that, generally, the error decreased with increased number of sensors, and that random sensor placement had negligible impact on errors as compared to using a sophisticated QR-based sensor placement scheme. However, as the surface-to-subsurface sensing problem studied in the current paper differs from what has been considered in the previous SHRED application, and the limiting case with 50 sensors in \cite{williams2024} still uses a low number of sensors compared to available data points, we include a brief investigation of the reconstruction performance dependency on the sensor count here.
}

\begin{figure}[!htbp]
    \centering
    
    \begin{subfigure}[b]{0.48\textwidth}
         \centering
         \includegraphics[width=\textwidth]{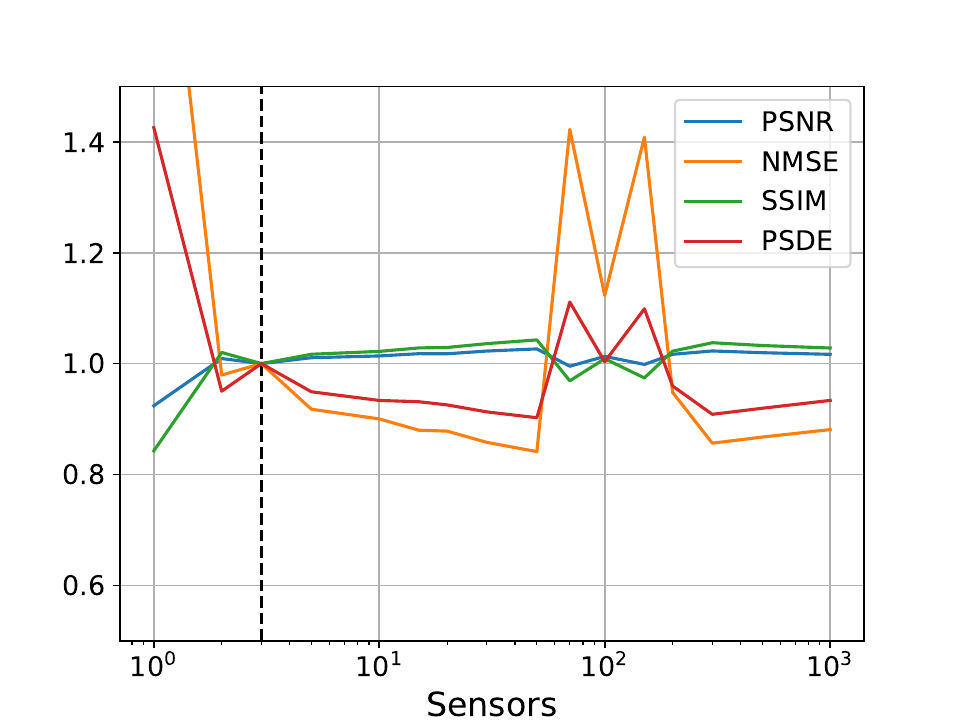}
         \caption{S1}
         
     \end{subfigure}
              \begin{subfigure}[b]{0.48\textwidth}
         \centering
         \includegraphics[width=\textwidth]{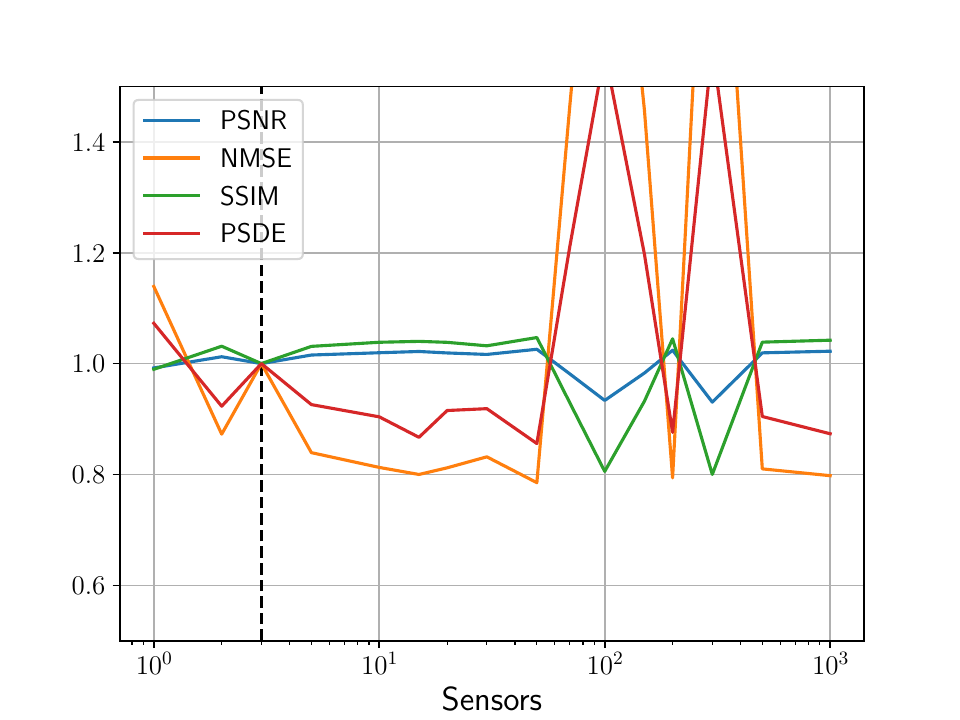}
         \caption{S2}
        
     \end{subfigure}
    \hfill
    \vspace{0.9cm}
     \hfill
     \begin{subfigure}[b]{0.48\textwidth}
         \centering
         \includegraphics[width=\textwidth]{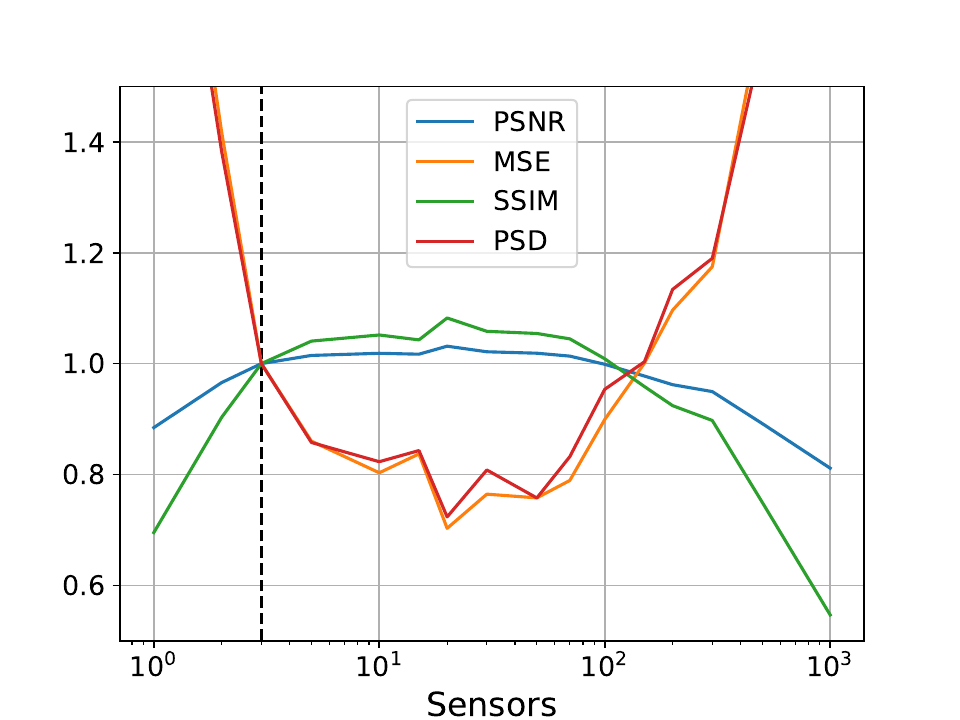}
         \caption{E1}
         
     \end{subfigure}
          \begin{subfigure}[b]{0.48\textwidth}
         \centering
         \includegraphics[width=\textwidth]{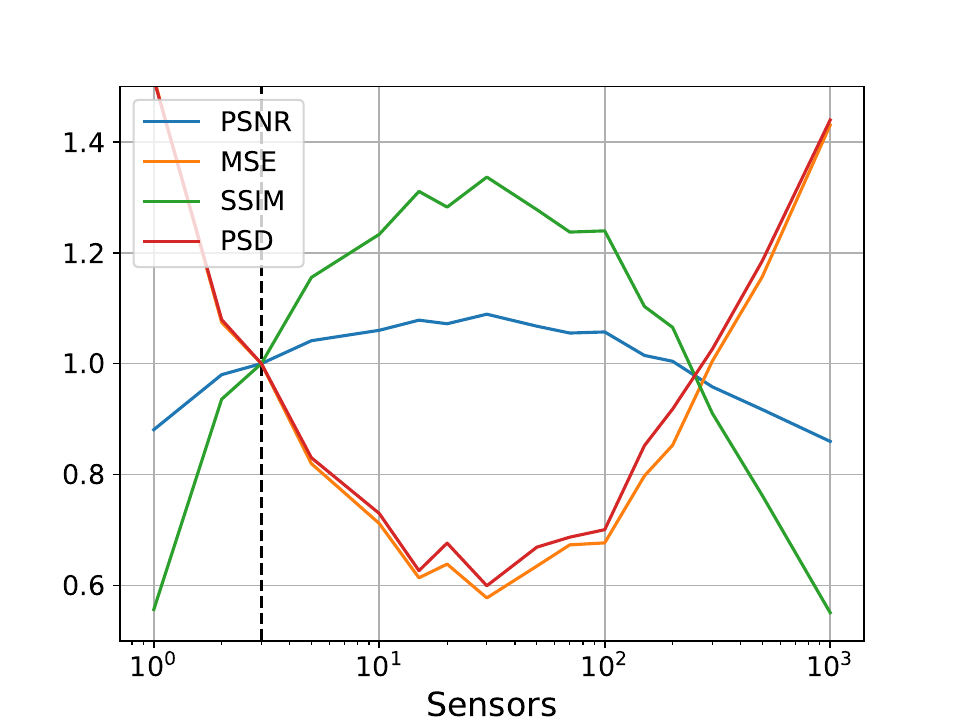}
         \caption{E2}
         
     \end{subfigure}
     \hfill
    \caption{\akr{Normalized error metrics averaged in depth for a range of sensor configurations. The dashed line marks the sensor configuration used throughout the paper, which is three randomly placed points sensors at the surface. The error metrics are normalized based on the mean error at this configuration. Note that optimal SHRED performance is met if PSNR and SSIM are maximal, while the same is true when PSD and MSE are minimal.}}
    \label{fig:sensor_params}
\end{figure}

\akr{
To investigate the impact the sensor count has on SHRED performance, we varied the number of sensors from 1 to 1000, and ran an ensemble of 20 trained SHRED models per sensor configuration. 
For each sensor configuration, we compute depth-averaged error metrics, in the same way as for the investigation of rank in \S \ref{sec:app1}1. The results are normalize by the results for the three-sensor configuration used throughout this paper.
Figure \ref{fig:sensor_params} depicts the results for all four datasets. We first observe that using an order of magnitude more sensors would have had a positive impact on the results, for all datasets. In particular, for the experimental datasets E1 and E2, increasing to 20--40 sensors appears to be a safe and beneficial choice, but going much beyond this, to, say, 100 sensors, leads to an increase in the error. For the DNS datasets, on the other hand, we observe that although using 30 sensors reduces the reconstruction errors, a further increase in the number of sensors may lead to instabilities. This is evident in the abrupt jump in NMSE and PSDE for two of the tested sensor configurations for datasets S1 and S2, both outside the test regime reported in \cite{williams2024}.
}

%

\end{document}